\documentclass[aps,pra,longbibliography,superscriptaddress,floatfix]{revtex4-1}
\usepackage{amsfonts}
\usepackage{graphicx}
\usepackage{epsfig}
\usepackage{dcolumn}
\usepackage{bm}
\usepackage{amsmath}
\usepackage{graphicx}
\usepackage[latin1]{inputenc}
\usepackage{ulem}
\usepackage{epstopdf}
\usepackage{subfigure}
\usepackage{color}
\usepackage{amsthm}
\usepackage{newlfont}
\usepackage{graphicx}
\usepackage{epstopdf}
\usepackage{appendix}
\usepackage[breaklinks=true]{hyperref}
\usepackage{breakcites}
\usepackage{textcomp}
\usepackage{appendix}
\usepackage{multirow}
\usepackage{color}
\usepackage{amssymb}
\usepackage{epsfig}
\usepackage{mathptmx}
\usepackage{bm}
\usepackage[american]{babel}
\usepackage{braket}
\begin{document}

\title{Nonlinearity and discreteness: solitons in lattices}
\author{ Boris A. Malomed }

\address{
Department of Physical Electronics. \\
School of Electrical Engineering, Faculty of Engineering, and Center for Light-Matter Interaction \\
Tel Aviv University,\\
Tel Aviv 69978, Israel}
\begin{abstract}
An overview is given of basic models combining discreteness in
their linear parts (i.e., the models are built as dynamical lattices) and
nonlinearity acting at sites of the lattices or between the sites. The
considered systems include the Toda and Frenkel-Kontorova lattices
(including their dissipative versions), as well as equations of the discrete
nonlinear Schr\"{o}dinger and Ablowitz-Ladik types, and their
combination in the form of the Salerno model. The interplay of
discreteness and nonlinearity gives rise to a variety of states, most
important ones being discrete solitons. Basic results for 1D and 2D
discrete solitons are collected in the
review, including 2D solitons with embedded vorticity, and some results
concerning mobility of discrete solitons.Main experimental findings are
overviewed too. Models of the semi-discrete type, and basic results for
solitons supported by them, are also considered, in a brief form. Perspectives
for the development of topics covered the review are discussed throughout the
text.
\end{abstract}
\maketitle

\section{Introduction: discretization of continuum models, and the continuum
limit of discrete ones}

Standard models of dynamical media are based on partial differential
equations, typical examples being the nonlinear Schr\"{o}dinger (NLS)
equation for the mean-field complex wave function $\psi \left(
x,y,z.t\right) $ in atomic Bose-Einstein condensates (BECs; in that case,
the NLS equation is usually called the Gross-Pitaevskii equation (GPE) \cite%
{Pethick}), and the NLS equation for the envelope amplitude of the
electromagnetic field in optical media \cite{Gadi}. In the scaled form, the
NLS equation is%
\begin{equation}
i\psi _{t}=-(1/2)\nabla ^{2}\psi +g|\psi |^{2}\psi +U\left( x,y,z\right)
\psi ,  \label{NLSE}
\end{equation}%
where $g=+1$ and $-1$ correspond to the self-defocusing and focusing signs
of the local cubic nonlinearity, and $U\left( x,y,z\right) $ is a real
external potential. In the application to optics, the evolution variable $t$
is replaced by coordinate $z$ in the propagation direction, while original $z
$ is replaced by the temporal variable, $\tau =t-z/V_{\mathrm{gr}}$, where $t
$ is time, and $V_{\mathrm{gr}}$\ is the group velocity of the carrier wave
\cite{KA}. In optics, the effective potential may be two-dimensional (2D), $%
-U(x,y)$ being a local variation of the refractive index in the transverse
plane.

In many cases the potential represents a spatially periodic pattern, such as
optical lattices (OLs) in BEC \cite{OL,Porter}, or photonic crystals which
steer the propagation of light waves \cite{PhotCryst} in optics:
\index{\textbf{\large L}!Lattices!optical}
\begin{equation}
U_{\mathrm{latt}}\left( x,y,z\right) =-\varepsilon \left[ \cos (2\pi
L/x)+\cos (2\pi L/y)+\cos (2\pi L/z)\right] ,  \label{UOL}
\end{equation}%
as well as its 2D and 1D reductions. A deep lattice potential, which
corresponds to large $\varepsilon $, splits the continuous wave function
into an array of ``droplets\textquotedblright\ trapped in local potential
wells, which are coupled by weak tunneling. Accordingly, in the framework of
the \textit{tight-binding approximation}, the NLS equation is replaced by a
discrete NLS (DNLS) equation, which was derived, in the 1D form, for arrays
of optical fibers \cite{ChristoJoseph,Silberberg,discr-review-0,discr-review}
and arrays of plasmonic nanowires \cite{Nicolae}, as well as for BEC loaded
in a deep OL trap \cite{Smerzi}:%
\begin{gather}
i%
\dot{\psi}_{l,m,n}=-(1/2)\left[ \left( u_{l+1,m,n}+\psi _{l-1,m,n}-2\psi
_{l,m,n}\right) +\left( \psi _{l,m+1,n}+\psi _{l,m-1,n}-2\psi
_{l,m,n}\right) \right.  \notag \\
\left. +\left( \psi _{l,m,n+1}+\psi _{l,m,n-1}-2\psi _{l,m,n}\right) \right]
+g\left\vert \psi _{l,m,n}\right\vert ^{2}\psi _{l,m.n},  \label{DNLSE}
\end{gather}%
where the set of integer indices, $\left( l,m,n\right) $, replaces
coordinates $\left( x,y,z\right) $. DNLS equation (\ref{DNLSE}) is often
reduced to 2D and 1D forms. While it includes the linear coupling between
the nearest neighbors, 1D lattices can be built in the form of zigzag
chains, making it relevant to add couplings between the next-nearest
neighbors \cite{NNN,ChongNNN}. 2D lattices with similar additional coupling
are known too \cite{NNN2D}.

As concerns the sign parameter, $g=\pm 1$, Eq. (\ref{DNLSE}) admits flipping
$+1\leftrightarrow -1$ by means of the \textit{staggering transformation} of
the discrete wave function:%
\begin{equation}
\psi _{l,m,n}(t)\equiv (-1)^{l+m+n}\exp \left( -6it\right) \tilde{\psi}%
_{l,m,n}^{\ast }(t)  \label{stagg}
\end{equation}%
where $\ast $ stands for the complex-conjugate expression, and (in the 2D
and 1D situations, $\exp \left( -6it\right) $ is replaced by $\exp \left(
-4it\right) $ and $\exp \left( -2it\right) $, respectively).

It is well known that the 2D and 3D continuous NLS equation (\ref{NLSE})
with the self-focusing nonlinearity, i.e. $g<0$, gives rise to the\textit{\
critical} and \textit{supercritical} \textit{collapse}, respectively, i.e.,
appearance of singular solutions in the form of infinitely narrow and
infinitely tall peaks, after a finite evolution time \cite{Gadi}.\ The
discreteness arrests the collapse, replacing it by a \textit{quasi-collapse}
\cite{Laedke}, when the width of the shrinking peak becomes comparable to
the spacing of the DNLS lattice.
\index{\textbf{\large C}!Collapse!quasi-}

The DNLS equation and its extensions constitute a class of models with a
large number of physical realizations, which have drawn much interest as
subjects of mathematical studies as well \cite{DNLS-book}. The class also
includes systems of coupled DNLS equations \cite{Angelis,Herring}.

The 1D continuous NLS equation without the external potential and with
either sign of the nonlinearity, $g$, is integrable by means of the
inverse-scattering transform \cite{Zakh,Segur,Calogero,Newell}, although it
is nonintegrable in the 2D and 3D geometries. On the contrary to that, the
1D DNLS equation is not integrable, i.e., the direct discretization destroys
the integrability \cite{Herbst,non-integrable}. However, the continuous NLS
equation\ admits another discretization in 1D, which leads to an integrable
discrete model, viz., the Ablowitz-Ladik (AL) equation \cite{AL}:
\begin{equation}
i%
\dot{\psi}_{n}=-\left( \psi _{n+1}+\psi _{n-1}\right) \left( 1+\mu
\left\vert \psi _{n}\right\vert ^{2}\right) ,  \label{ALproper}
\end{equation}%
where positive and negative values of the real nonlinearity coefficient, $%
\mu $, correspond to the self-focusing and defocusing, respectively.
Considerable interest was also drawn to the nonintegrable combination of the
AL and DNLS equations, in the form of the Salerno model (SM)\ \cite{SA},
with an additional onsite cubic term, different from the intersite one in (%
\ref{ALproper}):%
\begin{equation}
i\dot{\psi}_{n}=-\left( \psi _{n+1}+\psi _{n-1}\right) \left( 1+\mu
\left\vert \psi _{n}\right\vert ^{2}\right) -2\left\vert \psi
_{n}\right\vert ^{2}\psi _{n},  \label{SAmodel}
\end{equation}%
with the magnitude and sign of the onsite nonlinearity coefficient fixed by
means of the rescaling and staggering transformation, respectively. The SM
finds a physical realization in the context of the Bose-Hubbard model, i.e.,
BEC loaded in a deep OL, in the case when dependence of the intersite
hopping rate on populations of the sites is taken into regard \cite%
{BH,BH-review}.

While the above-mentioned DNLS, AL, and SM discrete systems are derived as
the discretization of continuous NLS equations, one can look at this
relation in the opposite direction: starting from discrete equations, one
can derive their \textit{continuum limit}. In particular, in the case of the
SM equation (\ref{SAmodel}), the continuum approximation is introduced by
replacing the intersite combination of the discrete fields by a truncated
Taylor's expansion,
\begin{equation}
\psi _{n}(t)\equiv e^{2it}\Psi (x,t),\qquad \Psi \left( x=n\pm 1,t\right)
\approx \Psi \left( x=n\right) \pm \Psi _{x}{\big |}_{x=n}+(1/2)\Psi _{xx}{%
\big |}_{x=n},  \label{psiPsi}
\end{equation}%
where $\Psi (x)$ is treated as a function of continuous coordinate $x$,
which coincides with $n$ when it takes integer values. The substitution of
this approximation in (\ref{SAmodel}) leads to a generalized (nonintegrable)
form of the 1D NLS equation \cite{Zaragoza}
\begin{equation}
i\Psi _{t}=-\left( 1+\mu \left\vert \Psi \right\vert ^{2}\right) \Psi
_{xx}-2\left( 1+\mu \right) \left\vert \Psi \right\vert ^{2}\Psi ,
\label{SAcont}
\end{equation}%
which goes over into the standard 1D\ NLS equation (\ref{NLSE}) with $g=+1$
and $U=0$ in the case of $\mu =0$.

The objective of this Chapter is to present an overview of basic discrete
nonlinear models and dynamical states produced by them, chiefly in the form
of bright solitons (self-trapped localized modes). Before proceeding to
models based on equations of the DNLS, AL, and SM types, simpler ones, which
were derived for chains of interacting particles, are considered in the next
section. The paradigmatic model of the latter type is provided by the 1D
Toda-lattice (TL) equation \cite{Toda}, written for coordinates $u_{n}(t)$
of particles with unit mass and exponential potential of interaction between
adjacent ones:%
\begin{equation}
\ddot{u}_{n}+\,e^{-\left( u_{n+1}-u_{n}\right) }-e^{-\left(
u_{n}-u_{n-1}\right) }=0.  \label{TLEu}
\end{equation}%
This equation can also be written for separations $r_{n}(t)\equiv
u_{n+1}(t)-u_{n}(t)$ between the particles:
\begin{equation}
\ddot{r}_{n}+\,e^{-r_{n+1}}+e^{-r_{n-1}}-\,2e^{-r_{n}}=0.  \label{TLE}
\end{equation}%
Equation (\ref{TLE}) is integrable \cite{Zakh}, its continuum limit being
the so-called ``bad'' Boussinesq equation \cite{Johnson}, which is formally
integrable too \footnote{%
``bad'' implies that (\ref{Bouss}) gives rise to an unstable dispersion
relation.} \cite{Calogero}:

\begin{equation}
r_{tt}-r_{xx}-(1/12)r_{xxxx}+\left( r^{2}\right) _{xx}=0  \label{Bouss}
\end{equation}

Another famous, although not integrable, model of a chain of
pairwise-interacting particles with coordinates $u_{n}(t)$, is the
Fermi-Pasta-Ulam (FPU) system \cite{FPU,FPU2}:%
\begin{equation}
\ddot{u}_{n}=\left( u_{n+1}+u_{n-1}-2u_{n}\right) \left[ 1+\alpha \left(
u_{n+1}-u_{n-1}\right) \right] ,  \label{fpu}
\end{equation}%
where $\alpha $ is a constant. This model was one of the first objects of
numerical simulations performed in the context of fundamental research (in
1953, published in 1955 \cite{FPU}, see also \cite{FPU3}). Later, it became
known that a very essential contribution to the original FPU work was made
by Mary Tsingou \cite{Tsingou}, therefore the model is also called the
FPU-Tsingou system.

The initial objective of the original numerical FPU-Tsingou experiment was
to observe the onset of ergodicity in the evolution governed by (\ref{fpu}).
A surprising result was that long simulations demonstrated a quasi-periodic
evolution, without manifestations of ergodicity (i.e. without statistically
uniform distribution of the energy between all degrees of freedom of the
lattice system). Eventually, this perplexing result was explained (in the
same paper \cite{soliton} by N. Zabusky and M. Kruskal which had introduced
word ``soliton") by the fact that the continuum limit of (\ref{fpu}) may be
reduced (for unidirectional propagation in excitations in the continuum
medium) to the Korteweg -- de Vries equation, which, being integrable, does
not feature ergodicity.
\index{\textbf{\large S}!Soliton}

The next section briefly addresses, in addition to the TL, more complex
models which combine the inter-particle interactions (taken in the linear
approximation, unlike the exponential terms in (\ref{TLE})), and onsite
potentials -- most typically, in the form of 
$U=\varepsilon \sum_{n}\left( 1-\cos u_{n}\right) ,$ 
with $\varepsilon >0$, which is the source of the nonlinearity in the
corresponding Frenkel-Kontorova (FK) model. It was originally introduced as
a model for dislocations in a crystalline lattice \cite{FrKo}, and has found
a large number of realizations in other physical settings \cite{Braun}

This Chapter also addresses, in a brief form, other models of nonlinear
discrete systems. These are discrete multidimensional models, semi-discrete
ones, and experimental realizations of discrete media and bright solitons in
them, chiefly in the realm of nonlinear optics. Dissipative discrete
nonlinear systems are partly addressed in this Chapter, as a systematic
consideration of dissipative discrete systems is a subject for a separate
review.

Because the length of the Chapter is limited, the presentation and
bibliography are not aimed to be comprehensive; rather, particular results
mentioned in sections following below are selected as examples which help to
understand general principles supported by a large body of theoretical and
experimental findings.

\section{Excitations in chains of interacting particles}

\subsection{The Toda lattice}

\index{\textbf{\large L}!Lattices!Toda}

The TL equation (\ref{TLEu}) is characterized, first of all, by its linear
spectrum. Looking for solutions to the linearized version of the equation in
the form of ``phonon modes'', i.e. plane waves with an infinitesimal
amplitude $u^{(0)}$, frequency $\chi $ and wavenumber $p$ (which is
constrained to the first Brillouin zone, $0<p<2\pi $),
\begin{equation}
\left( u_{n}\right) _{\mathrm{linearized}}=u^{(0)}\exp \left( ipn-i\chi
t\right) ,  \label{phonon}
\end{equation}%
it is easy to obtain the respective dispersion relation,%
\begin{equation}
\chi =\pm 2\sin \left( p/2\right) .  \label{TLdisp}
\end{equation}%
Further, (\ref{phonon}) produces phase velocities of the linear waves, $V_{%
\mathrm{ph}}=\chi /p$, which take values $\left\vert V_{\mathrm{ph}%
}\right\vert <1$.

Integrable equation (\ref{TLEu}) generates exact soliton solutions, which
were first found in the original work of Toda \cite{Toda}. The soliton
represents a lattice deformation traveling at constant velocity $c$:
\index{\textbf{\large S}!Soliton!discrete}
\begin{equation}
u_{n}=-\ln \left[ \xi ^{-2}-%
\frac{\xi ^{-2}-1}{1+\xi ^{2(n-ct)}}\right] ,\quad c=\pm \frac{\xi ^{-1}-\xi
}{\ln \left( \xi ^{-2}\right) },  \label{Tsol}
\end{equation}%
where $\xi $ is an arbitrary real parameter taking values $0<\xi <1$, the
respective interval of the inverse velocities being
\begin{equation}
0<|c|^{-1}<1.  \label{c}
\end{equation}%
Note that this interval has no overlap with the above-mentioned range of the
phase velocities of the linear modes, $\left\vert V_{\mathrm{ph}}\right\vert
<1$, in accordance with the well-known principle that solitons may exist in
\textit{bandgaps} of linear spectra, i.e., in regions where linear waves do
not exist.

Comparing values of the solution (\ref{Tsol}) at $n\rightarrow \pm \infty $,
one concludes that the soliton carries compression of the TL by a finite
amount, $\Delta u\equiv u_{n=+\infty }-u_{n=-\infty }=\ln \left( \xi
^{-2}\right) $, while a characteristic width of the soliton is $\Delta n\sim
1/\ln \left( \xi ^{-2}\right) $. Similar to other integrable systems \cite%
{Zakh,Segur,Calogero,Newell}, collisions between solitons do not affect
their shapes and velocities, leading solely to finite shifts of the
solitons' centers.

The limit of $\xi \rightarrow 0$ implies that the TL reduces to a chain of
hard particles, which interact when they collide. Accordingly, the soliton's
structure degenerates into a single fast moving particle, the propagation
being maintained by periodically occurring collisions, as a result of which
the moving particle comes to a halt, transferring its momentum to the
originally quiescent one. In the opposite limit, $\xi \rightarrow 1$,
soliton (\ref{Tsol}) becomes a very broad solution, traveling with the
minimum velocity, $c\rightarrow 1$. As mentioned above, no TL solitons
exists with velocities $|c|<1$.

Equation (\ref{TLEu}) conserves the total momentum, $P=\sum_{-\infty
}^{+\infty }\dot{u}_{n}$, and Hamiltonian (energy),%
\begin{equation}
H_{\mathrm{TL}}=\sum_{n=-\infty }^{+\infty }\left\{ \frac{1}{2}\dot{u}%
_{n}^{2}+\left[ e^{-\left( u_{n+1}-u_{n}\right) }-1\right] \right\} .
\label{ETL}
\end{equation}%
In fact, integrable equations, including (\ref{TLEu}), conserve an infinite
number of dynamical invariants, the momentum and energy being the
lowest-order ones in the infinite sequence \cite{Zakh}; however,
higher-order invariants do not have a straightforward physical
interpretation.

A realistic implementation of the TL includes friction forces with
coefficient $\alpha >0$, which should be compensated by an \textquotedblleft
ac\textquotedblright\ (time-periodic) driving force with amplitude $%
\varepsilon $ and frequency $\omega $ \cite{drivenTL,Jarmo,Rasmussen}. The
accordingly modified equation (\ref{TLEu}) is%
\begin{equation}
\ddot{u}_{n}+\,e^{-\left( u_{n+1}-u_{n}\right) }-e^{\left(
u_{n}-u_{n-1}\right) }=-\alpha \dot{u}_{n}+\varepsilon q_{n}\cos \left(
\omega t\right) .  \label{friction}
\end{equation}%
Here coefficients $q_{n}$ may be realized as charges of the particles, if
the drive is applied by an ac electric field. Nontrivial coupling of the
field to the TL dynamics is not possible if all the charges are identical,
i.e. $q_{n}\equiv 1$. Indeed, in the latter case one can trivially eliminate
the drive by defining $u_{n}(t)\equiv v_{n}(t)-\varepsilon \omega ^{-2}\cos
\left( \omega t\right) $, ending up with an equation (\ref{friction}) for $%
v_{n}(t)$ with no drive. The simplest nontrivial coupling is provided by
assuming $q_{n}=(-1)^{n}$, i.e. alternating positive and negative charges at
neighboring sites of the TL \cite{drivenTL}. A particular choice of the
periodic pattern for $q_{n}$ defines the respective size, $a$, of the cell
of the ac-driven TL (in particular, $q_{n}=(-1)^{n}$ corresponds to $a=2$).

The periodic passage of the soliton running through the lattice with
velocity $c$, i.e., with temporal period $T=a/c$, may provide compensation
of the friction losses if it resonates with the periodicity of the ac drive,
which defines the spectrum of \textit{resonant velocities} \cite{drivenTL},%
\begin{equation}
c_{N}=\pm \omega a/\left[ 2\pi \left( 1+2N\right) \right] ,  \label{cN}
\end{equation}%
where integer $N=0,1,2,...$ determines the order of the resonance \footnote{%
for velocities given by (\ref{cN}) with odd integer $1+2N$ replaced by an
even one, $2N$, with $N=1,2,...,$ the transfer of energy from the drive to
the moving soliton averages to zero}. Velocities $c_{N}$ are relevant if
they satisfy restriction $|c_{N}|>1$ (see (\ref{c})), which implies $\omega
>2\pi /a$.

The progressive motion of solitons is actually supported by the drive whose
strength, $\varepsilon $, exceeds a certain minimum (threshold) value, $%
\varepsilon _{\mathrm{thr}}$, which is roughly proportional to the friction
coefficient, $\alpha $ \cite{Jarmo}.

A specific class of dynamical chains with essentially nonlinear interaction
between adjacent particles of a finite size (spheres) represents models of
1D granular media, in which spheres interact when they come in touch. It was
demonstrated that such chains (in particular, those with the Hertz potential
of the contact interaction \cite{Hertz}) support self-trapped states in the
form of discrete breathers \cite{Daraio}.

\subsection{The Frenkel-Kontorova model and related systems}

\index{\textbf{\large L}!Lattices!Frenkel-Kontorova}

A paradigmatic example of lattices which combine interactions between
adjacent particles and the onsite potential acting on each particle is
provided by the FK model \cite{Braun}, which is the discretization of the
commonly known sine-Gordon (sG) equation. In 1D, the sG equation for real
wave field $u$ is \cite{Zakh,Segur,Calogero,Newell}%
\begin{equation}
u_{tt}-u_{xx}+\sin u=0.  \label{SG}
\end{equation}%
Elementary solutions to (\ref{SG}) are kinks, with topological charge $%
\sigma \equiv \left[ u\left( x=+\infty \right) -u\left( x=-\infty \right) %
\right] /\left( 2\pi \right) =\pm 1$:%
\begin{equation}
u_{\mathrm{kink}}=\arctan \left[ \exp \left( \sigma \left( x-ct\right) /%
\sqrt{1-c^{2}}\right) \right] ,  \label{kink}
\end{equation}%
with the velocity taking values $-1<c<+1$.

The discretization of (\ref{SG}) with stepsize $h$ implies defining
\begin{equation}
x\equiv hn,u\left( x=hn\right) \equiv u_{n},  \label{xn}
\end{equation}%
and the replacement of the second derivative by its finite-difference
counterpart:%
\begin{equation}
u_{xx}\rightarrow h^{-2}\left( u_{n+1}+u_{n-1}-2u_{n}\right) .
\label{second}
\end{equation}%
The result is the FK model, which also includes the local friction with
coefficient $\alpha \geq 0$, and an external force $f_{n}$, that may be
time-dependent:%
\begin{equation}
\ddot{u}_{n}-(1/h^{2})\left( u_{n+1}+u_{n-1}-2u_{n}\right) +\sin
u_{n}=-\alpha \dot{u}_{n}+f_{n}(t).  \label{FK1D}
\end{equation}%
The linearization of (\ref{FK1D}), with $\alpha =f_{n}=0$, for phonon modes (%
\ref{phonon}) gives rise to the following spectrum:%
\begin{equation}
\chi ^{2}=1+\left( 4/h^{2}\right) \sin ^{2}(p/2),  \label{SGdisp}
\end{equation}%
cf. its counterpart (\ref{TLdisp}) for the TL. The form of spectrum (\ref%
{SGdisp}) implies that localized oscillatory states may exists in the inner
and outer bandgaps, with frequencies $\chi ^{2}<1$ and $\chi ^{2}>1+4/h^{2}$%
, respectively.

In the connection to the linear spectrum, it is relevant to mention that
considerable interest was recently drawn to specially designed discrete
lattices whose spectrum includes a \textit{flatband}, i.e. a degenerate
branch of the $\chi (p)$ dependence in the form of $\chi =\mathrm{const}$,
as such systems admit the existence of localized discrete modes in the
absence of nonlinearity \cite{flat,flat2}. Effects of nonlinearity on
localized states in flatband systems have been investigated too \cite%
{flat2,Zegadlo}.

Generally similar to the discrete sG lattice governed by (\ref{FK1D}) are
models based on discretization of Klein-Gordon equations. Typically, they
feature the onsite cubic nonlinearity, the simplest model being \cite{phi4}%
\begin{equation}
\ddot{u}_{n}-(1/h^{2})\left( u_{n+1}+u_{n-1}-2u_{n}\right)
-u_{n}+u_{n}^{3}=0.  \label{DKG}
\end{equation}%
%
The spectrum of the linearization of Eq. (\ref{DKG}) is \textit{unstable},
with $\chi ^{2}=-1+\left( 4/h^{2}\right) \sin ^{2}(p/2)$ taking negative
values. However, kink solutions, which connect constant values $u_{n}=\pm 1$
at $n\rightarrow \pm \infty $, are stable, as the constant nonzero
background is stable against small perturbations. As concerns moving kinks,
it is possible to construct a discrete model with a specially designed
combination of nonlinear terms, which admits exact solutions for moving
kinks with particular values of the velocity \cite{exact-kink}.

Even in the case of $\alpha =f_{n}=0$, the discrete sG equation (\ref{FK1D}%
), unlike its continuum counterpart (\ref{SG}), is not integrable.
Therefore, in the absence of the friction, the single dynamical invariant of
(\ref{FK1D}) is the energy, provided that the driving force is
time-independent:%
\begin{equation}
E=\sum_{n=-\infty }^{+\infty }\left[ (1/2)\dot{u}_{n}^{2}+(1/2)h^{-2}\left(
u_{n+1}-u_{n}\right) ^{2}+\left( 1-\cos u_{n}\right) -f_{n}u_{n}\right] .
\label{ESG}
\end{equation}%
Note that, treating $u_{n}$ as per Eq. (\ref{xn}), and similarly defining $%
f_{n}\equiv f(x=hn)$, one can formally write the energy as in the continuum
setting, in which the discreteness is introduced by means of a lattice of
delta-functions with period $h$:%
\begin{equation}
E=\int_{-\infty }^{+\infty }\!dx\sum_{n=-\infty }^{+\infty }\delta \left(
x-hn\right) \left\{ \frac{\dot{u}_{n}^{2}}{2}+\frac{\left[ u(x+h)-u(x)\right]
^{2}}{2h^{2}}+\left( 1-\cos u\right) -f(x)u\right\} .  \label{delta}
\end{equation}

A fundamentally important concept in models of the FK type is the
Peierls-Nabarro (PN) potential \cite{Ishimori}. It is naturally defined in
the quasi-continuum approximation, which implies that the lattice's spacing
is much smaller than a characteristic size of the mode under the
consideration, i.e., $h\ll 1$. In this limit case, the mode may be
considered in the continuum form -- e.g., as $u_{\mathrm{kink}}(x-\xi )$,
with the central point, $\xi $, placed at an arbitrary position, and the PN
potential is defined as the total energy, given by (\ref{delta}), considered
as a function of $\xi $ \cite{KC}. Then, using identity%
\begin{equation}
\sum_{n=-\infty }^{+\infty }\delta \left( x-hn\right) \equiv \frac{1}{h}%
\sum_{m=-\infty }^{+\infty }\exp \left( i\frac{2\pi m}{h}x\right) ,
\label{identity}
\end{equation}%
one obtains, in the lowest approximation, which is determined by the lowest
harmonics in expression (\ref{identity}), with $m=\pm 1$, an exponentially
small but, nevertheless, relevant result:%
\begin{equation}
U_{\mathrm{PN}}(\xi )=\frac{U_{0}}{2}\cos \left( \frac{2\pi \xi }{h}\right)
,\quad U_{0}=\frac{\left( 4\pi /h\right) ^{2}}{\sinh \left( \pi
^{2}/h\right) }.  \label{UPN}
\end{equation}%
Thus, the broad quasi-continuum\ mode tends to have its center pinned at any
local minimum of the PN potential, $\xi =h\left( (1/2)+N\right) $, with
arbitrary integer $N$. The PN potential barrier, which separates neighboring
minima, and thus creates an obstacle for free motion of kinks, is $U_{0}$.
The PN barrier may be suppressed in FK lattices with a long-range intersite
interaction added to the linear coupling between the nearest neighbors \cite%
{Shpyrko}.

Unlike the TL solitons (\ref{Tsol}), which may only exist as moving states
with velocities $|c|>1$, the existence of quiescent FK kinks, pinned to
local potential minima, is not predicated on the presence of the driving
force. On the other hand, the motion of kinks is braked by friction, as well
as by radiative losses, i.e., emission of lattice \textquotedblleft
phonons\textquotedblright\ by a kink moving through the lattice, the latter
effect usually being much weaker than friction. As well as in the TL model,
the motion of kinks can be supported by the ac drive, $f_{n}=(-1)^{n}%
\varepsilon \cos \left( \omega t\right) $, at the same resonant velocities
as given by (\ref{cN}), with $a=2$ \cite{Bonilla}.

A relevant physical realization of the FK model is provided by an array of
coupled long Josephson junctions (JJs) \cite{JJ1,JJ2} (each junction is a
narrow dielectric layer separating two bulk superconductors \cite{Barone}).
An accurate model of the array is provided by Eq. (\ref{FK1D}), where $%
f_{n}\equiv f$ represents the bias current applied to each junction, while $%
\alpha $ is the coefficient of Ohmic loss. Especially interesting is this
version of the FK with periodic boundary conditions, which corresponds to
the circular JJ array built of $N$ junctions \cite{Cirillo,Strogatz}, as it
gives rise to resonant interaction between a kink (in terms of JJs, it is a
\textit{fluxon}, i.e., quantum of the magnetic flux), moving at velocity $c$
in the ring-shaped array, and phonon modes whose phase velocity $\chi /p$,
determined by dispersion relation (\ref{SGdisp}), may coincide with $c$. The
periodicity of the array imposes the \textquotedblleft
quantization\textquotedblright\ condition on wavenumber $p$ in (\ref{SGdisp}%
),
\begin{equation}
p=\left( 2\pi /hN\right) m,~m=1,2,3,...~.  \label{pm}
\end{equation}

\begin{figure}[tb]
\begin{center}
\includegraphics[width=7cm,scale=1]{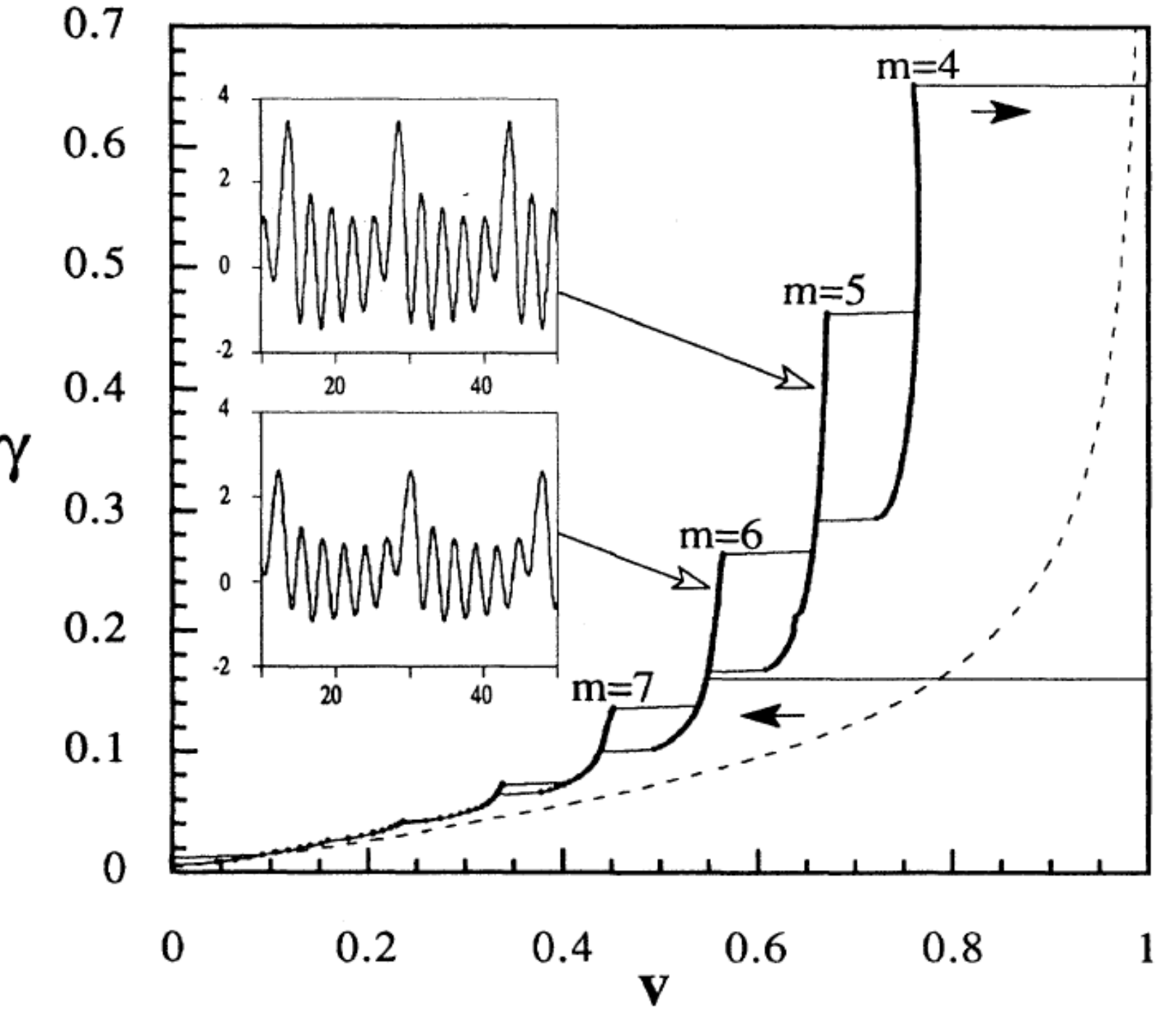}
\end{center}
\caption{The predicted relation between the velocity of the discrete kink,
travelling in the ring-shaped FK lattice composed of $N=10$ sites, subject
to periodic boundary conditions, and the driving force. Numbers $m$, which
label different vertical resonant steps, denote the \textquotedblleft
quantization\textquotedblright\ orders in Eq. (\protect\ref{pm}). Arrows
indicate directions of hysteretic jumps between the steps, and the dashed
curve represents the $f(c)$ dependence in the continuum version of the
system. Insets display dependences $\dot{u}_{n}$, which are proportional to
the local voltage in the underlying JJ array, at the system's midpoint,
corresponding to two points in the $f(c)$ (current-voltage) characteristic
marked by arrows. Parameters in Eq. (\protect\ref{FK1D}) are $h=1$ and $%
\protect\alpha =0.1$. Reproduced from \protect\cite{Cirillo}}
\label{fig1}
\end{figure}

The analysis of the kink-phonon interaction leads to a dependence of the
fluxon's velocity $c$ on the driving force (current), $f$, in the form of
resonant \textit{Shapiro steps} \cite{Shapiro} connected by hysteretic
jumps, as shown in Fig. \ref{fig1}. This dependence predicts an
experimentally observable current-voltage characteristic of the JJ system,
as the voltage is proportional to $c$. The measured characteristic was found
to be very close to the theoretical prediction \cite{Cirillo}.

Lastly, it is relevant to mention that the FK model also supports breathers,
i.e. localized modes which are periodically oscillating functions of time
\cite{Braun,Aubry1,Floria,Gendelman}. In the continuum limit, the breathers
naturally carry over into the well-known exact breather solutions of the sG
equation (\ref{SG}),

\section{Nonlinear Schr\"odinger (NLS) lattices}

\index{\textbf{\large L}!Lattices!Nonlinear Schr\"odinger}

\subsection{One-dimensional (1D) solitons}

\subsubsection{Fundamental states}

DNLS equation (\ref{DNLSE}) gives rise to discrete solitons, which cannot be
represented by analytical solutions, but can be easily found in a numerical
form. General properties of the soliton families can be understood by means
of the variational approximation (VA). Results for solitons in models of the
DNLS type are well known, being broadly represented in the literature \cite%
{DNLS-book}. Therefore, basic results for discrete NLS solitons are
summarized here in a brief form.
\index{\textbf{\large V}!Variational approximation}

Most studies addressed the 1D version of (\ref{DNLSE}), i.e.
\begin{equation}
i%
\dot{\psi}_{n}=-(1/2)\left( \psi _{n+1}+\psi _{n-1}-2\psi _{n}\right)
-\left\vert \psi _{n}\right\vert ^{2}\psi _{n},  \label{1D-DNLSE}
\end{equation}%
where the nonlinearity coefficient is fixed to be $g=-1$, which corresponds
to the self-focusing sign of the onsite nonlinearity (recall that the sign
of $g$ may be flipped by means of the staggering transformation (\ref{stagg}%
)). The DNLS equation conserves two dynamical invariants, \textit{viz}., the
total norm,%
\begin{equation}
N=\sum_{n=-\infty }^{+\infty }\left\vert \psi _{n}\right\vert ^{2},
\label{Ndiscr}
\end{equation}%
and Hamiltonian (energy),%
\begin{equation}
H=\sum_{n=-\infty }^{+\infty }\left[ (1/2)\left\vert \psi _{n}-\psi
_{n-1}\right\vert ^{2}-(1/4)\left\vert \psi _{n}\right\vert ^{4}\right] .
\label{H-DNLSE}
\end{equation}%
A fundamental property of the DNLS equation with the self-attractive onsite
nonlinearity is the modulational instability of its spatially homogeneous
state \cite{Peyrard}.

Stationary solutions to (\ref{1D-DNLSE}) with real frequency $\omega $
are looked for as
\begin{equation}
\psi _{n}(t)=e^{-i\omega t}u_{n},  \label{psi-u}
\end{equation}%
with real amplitudes $u_{n}$ satisfying the discrete equation,
\begin{equation}
\omega u_{n}=-(1/2)\left( u_{n+1}+u_{n-1}-2u_{n}\right) -u_{n}^{3}.
\label{1D-DNLS-stat}
\end{equation}%
While (\ref{1D-DNLS-stat}) does not admit exact analytical solutions, the VA
produces quite accurate approximations for discrete solitons. The VA is
based on the Lagrangian, from which (\ref{1D-DNLS-stat}) can be derived by
means of the variation with respect to discrete field $u_{n}$:%
\begin{equation}
L=\sum_{n=-\infty }^{+\infty }\left\{ (1/4)\left[ \left(
u_{n}-u_{n-1}\right) ^{2}-u_{n}^{4}\right] -\omega u_{n}^{2}\right\} .
\label{LagrDNLS}
\end{equation}

The use of the VA is based on a particular \textit{ansatz}, i.e. a trial
analytical expression which aims to approximate the solution \cite{Progress}%
. The only discrete ansatz for which analytical calculations are feasible is
represented by the exponential function \cite{Weinstein,Papa,Dave-VA,Gorder}%
. In particular, an \textit{onsite-centered }(OC) discrete soliton, i.e. one
with a single maximum (which is placed, by definition, at site $n=0$) is
approximated by%
\begin{equation}
\left( u_{n}\right) _{\mathrm{onsite}}=A\exp \left( -a|n|\right) ,
\label{ansatz-onsite}
\end{equation}%
with $a>0$. The corresponding norm, calculated as per (\ref{Ndiscr}), is%
\begin{equation*}
N_{\mathrm{ansatz}}=A^{2}\coth a\text{.}
\end{equation*}%
Note that ansatz (\ref{ansatz-onsite}) works well for strongly and
moderately discrete solitons (see Fig. \ref{fig2} below), but it is not
appropriate for broad (quasi-continuum) modes, which are approximated by the
commonly known soliton solution of the NLS equation (the 1D version of (\ref%
{NLSE}) with $U=0$),%
\begin{equation}
\psi \left( x,t\right) =\eta ~\mathrm{sech}\left( \eta \left( x-\xi \right)
\right) \exp \left( i\eta ^{2}t\right) ,  \label{cont-soliton}
\end{equation}%
with a large width, $\eta ^{-1}\gg 1$, and central coordinate $\xi $.

For \textit{intersite-centered} (IC) discrete solitons, with two symmetric
maxima placed at two adjacent sites of the lattice, $n=0$ and $n=1$ (and a
formal central point located between the sites, hence the name of these
modes), an appropriate ansatz is%
\begin{equation}
\left( u_{n}\right) _{\mathrm{intersite}}=A\exp \left( -a|n-1/2|\right) .
\label{ansatz-inter}
\end{equation}

The substitution of ansatz (\ref{ansatz-onsite}) in Lagrangian (\ref%
{LagrDNLS}) and straightforward calculations yield the following effective
Lagrangian:%
\begin{equation}
L_{\mathrm{eff}}=(A^{2}/2)\coth (a/2)-(A^{4}/4)\coth \left( 2a\right)
-\omega A^{2}\coth a.  \label{Leff}
\end{equation}%
Then, for given $\omega <0$ (the solitons with $\omega >0$ do not exist),
the squared amplitude, $A^{2}$, and inverse width, $a$, of the discrete
soliton are predicted by the Euler-Lagrange equations,%
\begin{equation}
\frac{\partial L_{\mathrm{eff}}}{\partial \left( A^{2}\right) }=\frac{%
\partial L_{\mathrm{eff}}}{\partial a}=0.  \label{EulerLagr}
\end{equation}%
This corresponding system of algebraic equations for $A^{2}$ and $a$ can be
easily solved numerically. A similar analysis was performed for the IC
solitons, starting with ansatz (\ref{ansatz-inter}). The VA produces quite
accurate predictions for solitons of both types, see Fig.~\ref{fig2} and
Ref.~\cite{Needs07}.

\begin{figure}[t]
\caption{Comparison of a typical OC (left panel) and IC (right panel) 1D
discrete solitons, obtained as numerical solutions of (\protect\ref%
{1D-DNLS-stat}), shown by chains of blue dots, and their counterparts
produced by the VA (shown by red open circles). In this figure, $\protect%
\omega =-1$, see Eq. (\protect\ref{psi-u}).}
\label{fig2}\includegraphics[width=7.5cm]{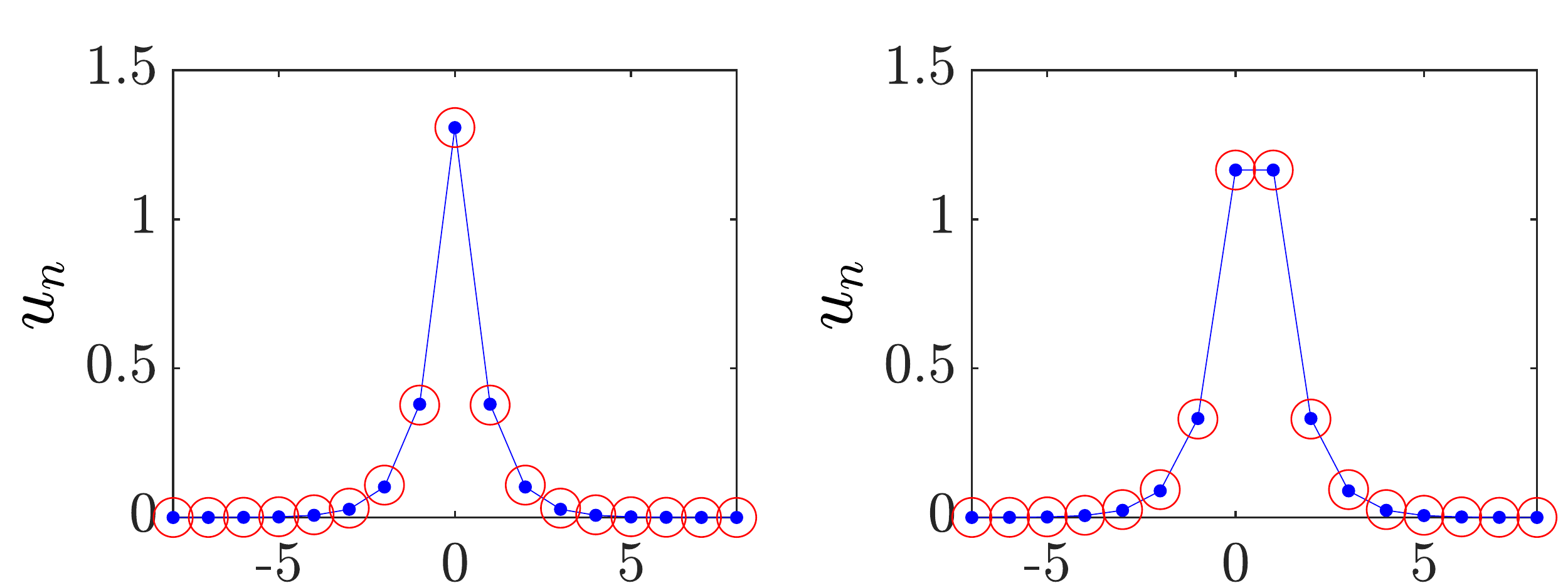} 
\end{figure}

An extended version of VA for 1D discrete solitons was elaborated for
nonstationary solutions, and compared to their numerically generated
counterparts \cite{Weinstein,Dave-VA}. Moreover, it was demonstrated that VA
may be applied, in a more sophisticated form, even to a challenging problem
of collisions of moving discrete solitons \cite{Papa}. Further
considerations addressed false instabilities, which are sometimes predicted
by the nonstationary VA \cite{Hadi-false instability}, and rigorous
justification of the VA \cite{rigorous}. Finally, the VA and full numerical
considerations demonstrate that the \emph{entire family} of the OC discrete
solitons is \emph{stable}, while all the IC ones are \emph{unstable} \cite%
{DNLS-book}. 

\subsubsection{Mobility of 1D discrete solitons}

\index{\textbf{\large S}!Soliton!Mobility}

The DNLS equation does not admit solutions for moving discrete solitons.
Indeed, even in the quasi-continuum approximation, soliton (\ref%
{cont-soliton}) is running through the effective PN potential, which, for 1D
DNLS modes, is%
\begin{equation}
U_{\mathrm{PN}}(\xi )=-%
\frac{\pi ^{4}}{3\sinh \left( \pi ^{2}/\eta \right) }\cos \left( 2\pi \xi
\right) ,  \label{UPN-NLS}
\end{equation}%
cf. expression (\ref{UPN}) for the PN barrier in the FK model. The periodic
acceleration and deceleration of the quasi-continuous soliton moving across
the PN potential gives rise to emission of small-amplitude \textquotedblleft
phonon" waves, i.e., losses which brake the motion. However, the emission
effect is extremely weak in direct simulations of the DNLS equations,
allowing the 1D discrete solitons to run\ indefinitely long \cite{Feddersen}%
. On the other hand, discrete solitons in the 2D DNLS equation (see the
following subsection) have no mobility. This is explained by the fact that
the above-mentioned quasi-collapse effect \cite{Laedke} makes them very
narrow modes strongly pinned to the underlying lattice.

The mobility of 1D discrete solitons in NLS lattices may be essentially
enhanced by means of the \textit{nonlinearity management }technique\textit{\
}\cite{management}, i.e., replacing coefficient $g$ in the 1D version of (%
\ref{DNLSE}) by a combination of constant (\textquotedblleft dc") and
time-periodic (\textquotedblleft ac") terms \cite{JC1}:
\begin{equation}
i\dot{u}_{n}+u_{n+1}+u_{n-1}-2u_{n}+\left[ g_{\mathrm{dc}}+g_{\mathrm{ac}%
}\sin (\omega t)\right] |u_{n}|^{2}u_{n}=0.  \label{dyn}
\end{equation}%
Similar to the situation for the damped driven TL, outlined above, discrete
solitons may move across the lattice at special values of the velocity,
determined by the resonance between the periodic passage of lattice sites by
the soliton and periodically oscillating ac component of the nonlinearity
coefficient in (\ref{dyn}), cf. (\ref{cN}):
\begin{equation}
\left( c_{\mathrm{res}}\right) _{N}^{(M)}=\frac{M\omega }{2\pi N},
\label{cres}
\end{equation}%
where integers $N$ and $M$ determine the order of the resonance. This
prediction was corroborated by simulations of (\ref{dyn}) \cite{JC1}.

\subsubsection{Higher-order modes in the 1D DNLS equation: twisted solitons
and bound states}

In addition to the OC and IC solitons, which are fundamental states, Eq. (%
\ref{1D-DNLS-stat}) admits stable higher-order states in the form of \textit{%
twisted modes}, which are subject to the antisymmetry condition, $%
u_{n}=-u_{1-n}$ \cite{twisted}. Such states exist and are stable only in a
strongly discrete form, vanishing in the continuum limit.

Stable discrete NLS solitons of the OC type may form bound states, which
also represent higher-order modes of the DNLS equation. They are stable in
the \textit{out-of-phase} form, i.e., for opposite signs of the bound
solitons \cite{bound states,bound states 2} (the same is true for 2D
discrete solitons \cite{Bishop}). Stationary bound states do not exist
either in the continuum limit, where bound states of NLS solitons are
represented solely by periodically oscillating \textit{breathers} \cite%
{Satsuma Yajima}.

\subsection{Two-dimensional (2D) discrete solitons and solitary vortices in
quiescent and rotating lattices}

\index{\textbf{\large V}!Vortices}
\index{\textbf{\large S}!Solitons!discrete}

\subsubsection{Static lattices}

The 2D cubic DNLS equation is a straightforward extension of the 1D equation
(\ref{1D-DNLSE}). In particular, its stationary form is
\begin{equation}
\omega u_{n}=-(1/2)\left(
u_{m+1,n}+u_{m-1,n}+u_{m,n+1}+u_{m,n-1}-4u_{m,n}\right) -\left\vert
u_{m,n}\right\vert ^{2}u_{m,n},  \label{2D-DNLSE}
\end{equation}%
cf. (\ref{1D-DNLS-stat}), where the stationary discrete wave function, $%
u_{m,n}$, may be complex. Fundamental-soliton solutions to (\ref{2D-DNLSE})
can also be predicted by means of VA \cite{Weinstein-2D,Chong} (see (\ref%
{SSB-ansatz}) below for the simplest 2D ansatz). More interesting in the 2D
case are discrete solitons with \textit{embedded vorticity}, which were
introduced in \cite{we} (see also \cite{they}). Vorticity, alias topological
charge, is defined as $\Delta \varphi /\left( 2\pi \right) $, where $\Delta
\varphi $ is a change of the phase of complex discrete wave function $%
u_{m,n} $, corresponding to any contour surrounding the vortex' pivot.
Stability is an important issue for 2D discrete solitons, because it is
commonly known that, in the continuum limit, the NLS equation in 2D gives
rise solely to unstable solitons, including fundamental ones (usually called
Townes' solitons \cite{Townes}), which are unstable against the critical
collapse \cite{Gadi}, and solitons with embedded vorticity \cite{Minsk},
which are still more unstable \cite{PhysD}.
\index{\textbf{\large C}!Collapse}

A typical example of a stable 2D discrete soliton is displayed in Fig. \ref%
{fig4}. 2D fundamental and vortex solitons, with topological charges $S=0$
and $1$, remain stable at $-\omega >|\omega _{\mathrm{cr}}^{(S=0)}|\approx
0.50$ \ and $-\omega >|\omega _{\mathrm{cr}}^{(S=1)}|\approx 1.23$,
respectively \cite{we}, while the higher-order localized discrete vortices
with $S=2$ and $4$ are unstable, being replaced by stable modes in the form
of quadrupoles and octupoles \cite{Zhigang}. Higher-order vortex solitons
with $S=3$ are stable only in a strongly discrete form, at $-\omega >|\omega
_{\mathrm{cr}}^{(S=2)}|\approx 4.94$.
\begin{figure}[t]
\caption{A stable discrete vortex soliton with topological charge $S=1$,
produced by Eq. (\protect\ref{2D-DNLSE}). The left and right panels show,
respectively, distributions of the absolute value and phase of complex wave
function $u_{m,n}$ in the plane of $\left( m,n\right) $.}
\label{fig4}
\includegraphics[width=7.5cm,scale=1]{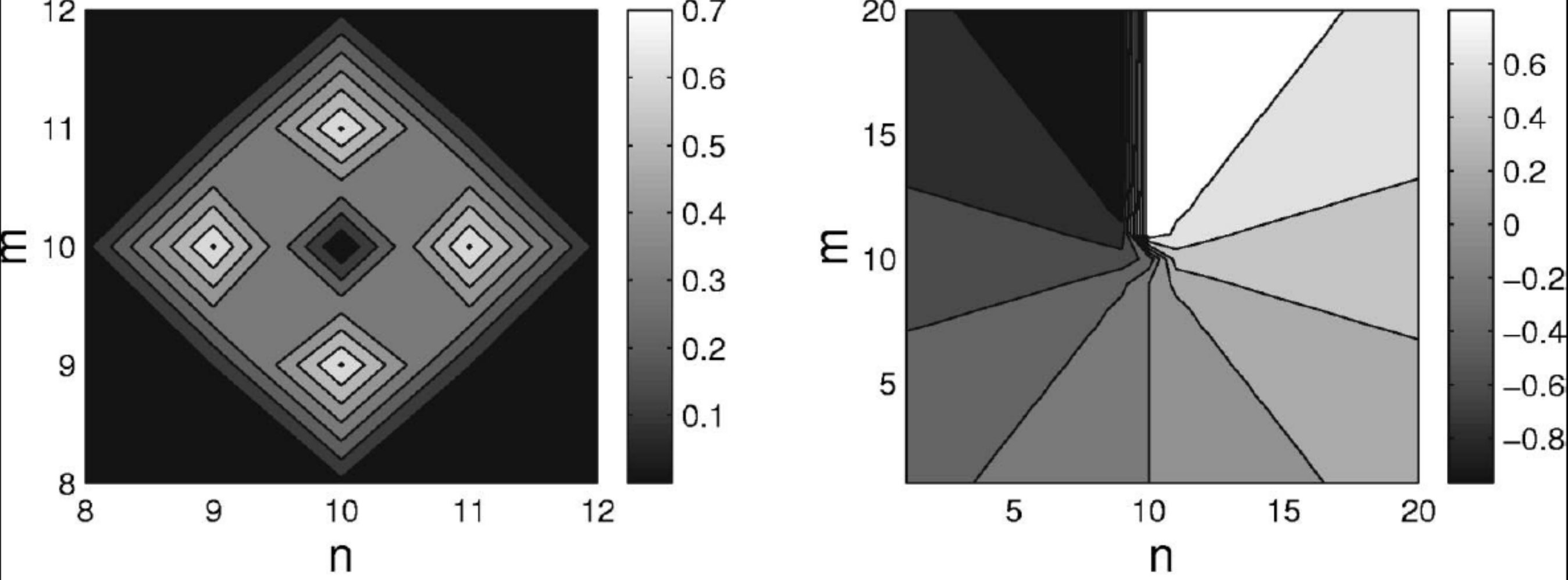}
\end{figure}

\begin{figure}[b]
\includegraphics[width=7.5cm,scale=1]{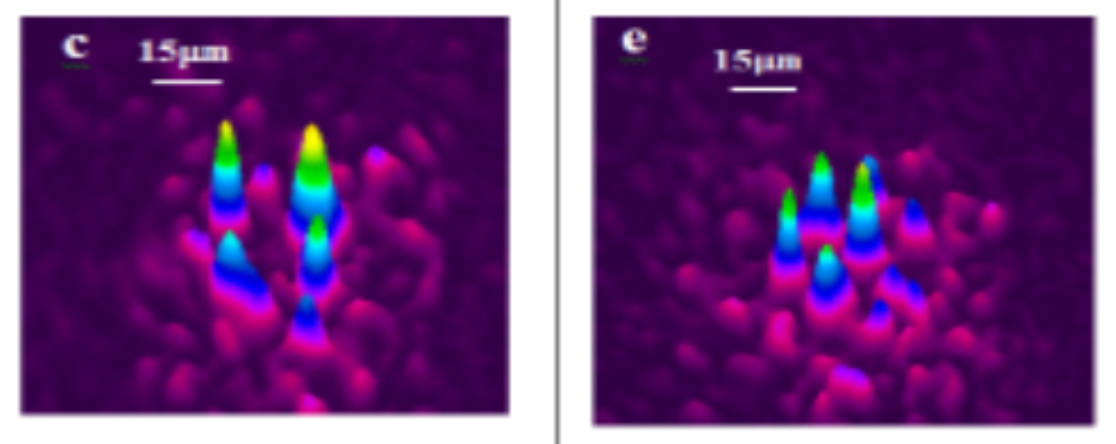}
\caption{Quasi-discrete optical solitons with vorticity $S=1$, created in a
bulk photorefractive crystal with an induced deep photonic lattice. The left
and right panels display, respectively, OC and IC vortex solitons.
Reproduced from \protect\cite{Segev}.}
\label{fig5}
\end{figure}

The theoretically predicted 2D discrete solitons with vorticity $S=1$ were
experimentally created in \cite{Kivshar} and \cite{Segev}, using a
photorefractive crystal. Unlike uniform media of this type, where
delocalized (``dark") optical vortices were originally produced \cite%
{Zhig1,Zhig2}, these works made use of a deep virtual photonic lattice as a
quasi-discrete medium supporting nonlinear optical modes in light beams with
extraordinary polarization (while the photonic lattice was induced by the
interference of quasi-linear beams in the ordinary polarization). Intensity
distributions observed in vortex solitons of the OC and IC types are
displayed in Fig. \ref{fig5}.

Another interesting result demonstrated (and theoretically explained) in
deep virtual photonic lattices is a possibility of periodic flipping of the
topological charge of a vortex soliton initially created with $S=2$ \cite%
{Chen}.

\subsubsection{Rotating lattices}

Dynamics of BEC loaded in OLs rotating at angular velocity $\Omega $, as
well as the propagation of light in a twisted nonlinear photonic crystal
with pitch $\Omega $, is modeled by the 2D version of Eq. (\ref{NLSE}),
written in the rotating reference frame:
\begin{equation}
i\psi _{t}=-\left( (1/2)\nabla ^{2}+\Omega
\hat{L}_{z}\right) \psi -\varepsilon \left[ \cos \left( kx\right) +\cos (ky)%
\right] \psi +g|\psi |^{2}\psi ,  \label{rotating}
\end{equation}%
where $\hat{L}_{z}=i(x\partial _{y}-y\partial _{x})\equiv i\partial _{\theta
}$ is the operator of the $z$-component of the orbital momentum ($\theta $
is the angular coordinate in the $\left( x,y\right) $ plane). In the
tight-binding approximation, Eq. (\ref{rotating}) is replaced by the
following variant of the DNLS equation \cite{JC2}:
\begin{gather}
i\dot{\psi}_{m,n}=-(C/2)\left\{ \left( \psi _{m+1,n}+\psi _{m-1,n}+\psi
_{m,n+1}+\psi _{m,n-1}-4\psi _{m,n}\right) \right.  \notag \\
\left. -i\Omega \left[ m\left( \psi _{m,n+1}-\psi _{m,n-1}\right) -n\left(
\psi _{m+1,n}-\psi _{m-1,n}\right) \right] \right\} +g|\psi _{m,n}|^{2}\psi
_{m,n}~,  \label{discrete}
\end{gather}%
where $C$ is the intersite coupling constant. In \cite{JC2}, stationary
solutions to (\ref{discrete}) were looked for in the form of ansatz (\ref%
{psi-u}), fixing $\omega \equiv -1$ and varying $C$ in (\ref{discrete}) as a
control parameter. Two species of localized states were thus constructed:
off-axis fundamental discrete solitons, placed at distance $R$ from the
rotation pivot, and on-axis ($R=0$) vortex solitons, with vorticities $S=1$
and $2$. At a fixed value of rotation frequency $\Omega $, a stability
interval for the fundamental soliton, $0<C<C_{\max }^{\mathrm{(fund)}}(R)$,
monotonously shrinks with the increase of $R$, i.e., most stable are the
discrete solitons with the center placed at the rotation pivot. Vortices
with $S=1$ are gradually destabilized with the increase of $\Omega $ (i.e.,
their stability interval, $0<C<C_{\max }^{\mathrm{(vort)}}(\Omega )$,
shrinks). On the contrary, a remarkable finding is that vortex solitons with
$S=2$, which, as said above, are completely unstable in the usual DNLS
equation with $\Omega =0$, are \emph{stabilized} by the rotation, in an
interval $0<C<C_{\mathrm{cr}}^{(S=2)}(\Omega )$, with $C_{\mathrm{cr}%
}^{(S=2)}(\Omega )$ growing as a function of $\Omega $. In particular, $C_{%
\mathrm{cr}}^{(S=2)}(\Omega )\approx 2.5\Omega $ at small $\Omega $ \cite%
{JC2}.

\subsection{Spontaneous symmetry breaking in linearly-coupled lattices}

\index{\textbf{\large S}!Spontaneous symmetry breaking}

A characteristic feature of many nonlinear \textit{dual-core} systems, built
of two identical linearly-coupled waveguides with intrinsic nonlinearity, is
a \textit{spontaneous-symmetry-breaking }(SSB)\textit{\ bifurcation}, which
destabilizes the symmetric ground state, with equal components in the
coupled cores, and creates stable asymmetric ones, when the nonlinearity
strength exceeds a critical value \cite{SSB}. A system of linearly-coupled
DNLS equations is a basic model for SSB in discrete settings. Its 2D form is
\cite{herring}
\begin{eqnarray}
i%
\dot{\phi}_{n} &=&-(1/4)\left( \phi _{m+1,n}+\phi _{m-1,n}+\phi
_{m,n+1}+\phi _{m,n-1}-4\phi _{m,n}\right) -\left\vert \phi
_{m,n}\right\vert ^{2}\phi _{m,n}-K\psi _{m,n}~,  \notag \\
i\dot{\psi}_{n} &=&-(1/4)\left( \psi _{m+1,n}+\psi _{m-1,n}+\psi
_{m,n+1}+\psi _{m,n-1}-4\psi _{m,n}\right) -\left\vert \psi
_{m,n}\right\vert ^{2}\psi _{m,n}-K\phi _{m,n}~,  \label{coupled}
\end{eqnarray}%
where $\phi _{m,n}$ and $\psi _{m,n}$ are discrete fields,\ and $K>0$
accounts for the linear coupling between them. Stationary states are looked
for as $\left( \phi _{m,n},\psi _{m,n}\right) =\exp \left( -i\omega t\right)
\left( u_{m,n},v_{m.n}\right) $, where the linear coupling makes it
necessary to have identical frequencies, $\omega $, in both components. Real
stationary fields are characterized by their norms,%
\begin{equation}
E_{u,v}=\sum_{m,n=-\infty }^{+\infty }\left( u_{m,n}^{2},v_{m,n}^{2}\right) ,
\label{E}
\end{equation}%
which define the asymmetry degree of the symmetry-broken states:%
\begin{equation}
r=\left( E_{u}-E_{v}\right) /\left( E_{u}+E_{v}\right) .  \label{r}
\end{equation}

\begin{figure}[tb]
\begin{center}
\begin{tabular}{cc}
\includegraphics[width=5.5cm,scale=1]{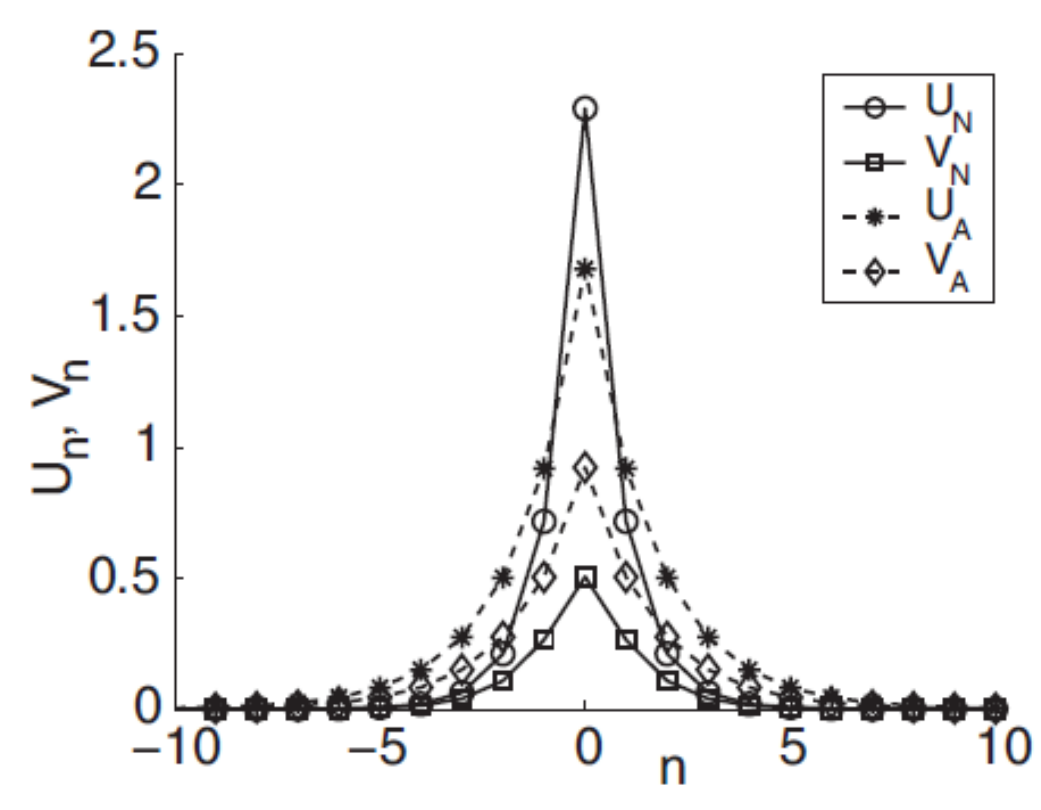} & %
\includegraphics[width=5.5cm,scale=1]{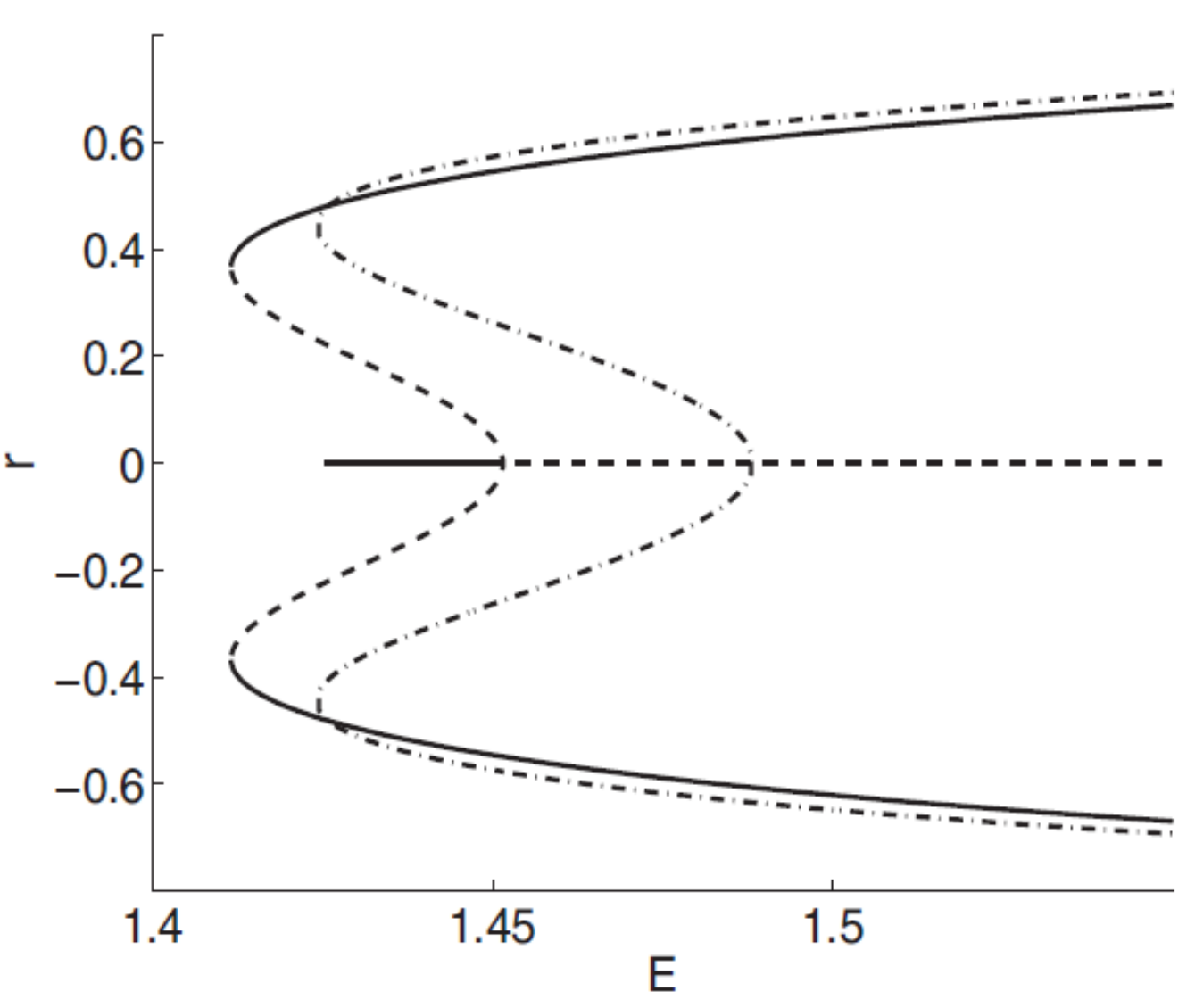}%
\end{tabular}%
\end{center}
\caption{(a) A stable 2D two-component discrete soliton with spontaneously
broken symmetry between the components, generated by system (\protect\ref%
{coupled}). The 2D soliton, with total norm $E\equiv E_{u}+E_{v}=1.435$, is
displayed by means of its 1D cross section. Symbols labelled $\left( U_{%
\mathrm{N}},V_{\mathrm{N}}\right) $ and $\left( U_{\mathrm{A}},V_{\mathrm{A}%
}\right) $ stand, respectively, for the components of the numerically
constructed soliton and its analytical counterpart predicted by the VA based
on ansatz (\protect\ref{SSB-ansatz}). (b) Families of 2D discrete solitons
of the OC type, generated by system (\protect\ref{coupled}), are shown by
means of curves $r(E)$ (where $r$ is the asymmetry parameter (\protect\ref{r}%
)). The dashed-dotted curve shows the VA prediction, while the solid and
dashed ones depict stable and unstable solitons produced by the numerical
solution. Reproduced from \protect\cite{herring}.}
\label{fig6}
\end{figure}

The present system can be analyzed by means of the VA, which is based on the
simplest ansatz (cf. its 1D counterpart (\ref{ansatz-onsite})):%
\begin{equation}
\left( u_{m,n},v_{m,n}\right) =\left( A,B\right) \exp \left[ -a\left(
|m|+|n|\right) \right] ,  \label{SSB-ansatz}
\end{equation}%
with inverse width $a$ and amplitudes, $A$ and $B$, of the two components.
The ansatz accounts for the SSB in the case of $A\neq B$. A typical example
of a \emph{stable} 2D discrete OC soliton is displayed in Fig. \ref{fig6}a,
which corroborates accuracy of the VA. The full set of symmetric and
asymmetric 2D discrete solitons is characterized, in Fig. \ref{fig6}b, by
the dependence of asymmetry parameter $r$, defined in (\ref{r}), on the
total norm, $E\equiv E_{u}+E_{v}$. It is seen that the SSB bifurcation is
one of a clearly \textit{subcritical} type \cite{Iooss}, with the two
branches of broken-symmetry states originally going backward as unstable
ones, and getting stable after passing the turning point. Accordingly,\ Fig. %
\ref{fig6}b demonstrates a considerable bistability area, where symmetric
and asymmetric states coexist as ones stable against small perturbations.


\section{Ablowitz-Ladik and Salerno-model lattices}

\subsection{1D models}

\index{\textbf{\large L}!Lattices!Ablowitz-Ladik}
\index{\textbf{\large S}!Salerno model}

1D\ models of AL and SM types, which are defined by Eqs. (\ref{ALproper})
and (\ref{SAmodel}), conserve the total norm, but its definition is
different from the straightforward one, given by Eq. (\ref{Ndiscr}) for the
DNLS equation; namely,
\begin{equation}
N_{\mathrm{AL,SM}}=(1/\mu )\sum_{n}\ln \left\vert 1+\mu |\psi
_{n}|^{2}\right\vert  \label{eq:Norm}
\end{equation}%
\cite{AL,Cai}. The Hamiltonian of the AL and SM equations is also
essentially different from the ``naive\textquotedblright\ DNLS Hamiltonian
given by Eq. (\ref{H-DNLSE}). As found in the original work of Ablowitz and
Ladik, the Hamiltonian of their model is%
\begin{equation}
H_{\mathrm{AL}}=-\sum_{n}\left( \psi _{n}\psi _{n+1}^{\ast }+\psi _{n+1}\psi
_{n}^{\ast }\right) ,  \label{HAL}
\end{equation}%
while for the SM, it is \cite{Cai}%
\begin{equation}
H_{\mathrm{SM}}=-\sum_{n}\left[ \left( \psi _{n}\psi _{n+1}^{\ast }+\psi
_{n+1}\psi _{n}^{\ast }\right) +(2/\mu )|\psi _{n}|^{2}\right] +(2/\mu )N_{%
\mathrm{AL}}.  \label{HSM}
\end{equation}%
The price paid for ostensible \textquotedblleft
simplicity\textquotedblright\ of expression (\ref{HAL}) is the complex form
of the respective Poisson brackets, which determine the dynamical equations
in terms of the Hamiltonian as $d\psi _{n}/dt=\left\{ H,\psi _{n}\right\} $.
For the AL and SM models, the Poisson brackets, written for a pair of
arbitrary functions of the discrete field variables, $B\left( \psi _{n},\psi
_{n}^{\ast }\right) ,$ $C\left( \psi _{n},\psi _{n}^{\ast }\right) $, are
\begin{equation}
\left\{ B,C\right\} =i\sum_{n}\left(
\frac{\partial B}{\partial \psi _{n}}\frac{\partial C}{\partial \psi
_{n}^{\ast }}-\frac{\partial B}{\partial \psi _{n}^{\ast }}\frac{\partial C}{%
\partial \psi _{n}}\right) \left( 1+\mu \left\vert \psi _{n}\right\vert
^{2}\right) .  \label{Poisson}
\end{equation}

As mentioned above, the continuum limit of the SM is represented by Eq. (\ref%
{SAcont}) \cite{Zaragoza}. This continuous equation conserves the total norm
and Hamiltonian, written in terms of variables $\Psi \left( x\right) $ (see
Eq. (\ref{psiPsi})), which are straightforward continuum counterparts of
expressions (\ref{eq:Norm})-(\ref{HSM}):
\begin{eqnarray}
\left( N_{\mathrm{AL}}\right) _{\mathrm{cont}} &=&\frac{1}{\mu }%
\int_{-\infty }^{+\infty }dx~{\ln }\left\vert 1+\mu |\Psi |^{2}\right\vert ,
\label{NALcont} \\
\left( H_{\mathrm{SM}}\right) _{\mathrm{cont}} &=&\int_{-\infty }^{+\infty
}dx\left[ \left\vert \Psi _{x}\right\vert ^{2}-2\left( \frac{1}{\mu }%
+1\right) |\Psi |^{2}\right] +\frac{2}{\mu }\left( N_{\mathrm{AL}}\right) _{%
\mathrm{cont}}  \label{HSMcont}
\end{eqnarray}

\subsection{Discrete 1D solitons}

\index{\textbf{\large I}!Integrability}

The AL equation (\ref{ALproper}) gives rise to an exact solution for
solitons in the case of self-focusing nonlinearity, $\mu >0$. Setting $\mu
\equiv +1$ by means of rescaling, the solution is%
\begin{equation}
\psi _{n}(t)=\left( \sinh \beta \right) \mathrm{sech}\left[ \beta (n-\xi (t))%
\right] \exp \left[ i\alpha \left( n-\xi (t)\right) -i\varphi (t)\right] ,
\label{ALsoliton}
\end{equation}%
where $\beta $ and $\alpha $ are arbitrary real parameters that determine
the soliton's amplitude, $A\equiv \sinh \beta $, its velocity,
$V\equiv
\dot{\xi}=2\beta ^{-1}\left( \sinh \beta \right) \sin \alpha $, 
and overall frequency 
$\Omega \equiv \dot{\varphi}=-2\left[ \left( \cosh \beta \right) \cos \alpha
+(\alpha /\beta )\left( \sinh \beta \right) \sin \alpha \right] $%
.

The existence of exact solutions for traveling solitons in the discrete
system is a highly nontrivial property of the AL equation, which follows
from its integrability. If the system is not integrable, motion of a
discrete soliton through a lattice is hampered by emission of radiation,
even if this effect may seem very weak in direct simulations \cite{Feddersen}%
. On the other hand, there are some special discrete equations which are not
integrable, but admit particular solutions for traveling solitons (at
exceptional values of the velocity, rather than at an arbitrary velocities,
as in the case of the AL solitons \cite{Kevrekidis,Oxtoby}.

The stationary version of the SM, obtained by the substitution of the usual
ansatz (\ref{psi-u}), with real $u_{n}$, in (\ref{SAmodel}), is%
\begin{equation}
\omega u_{n}=-\left( u_{n+1}+u_{n-1}\right) \left( 1+\mu u_{n}^{2}\right)
-2u_{n}^{3},  \label{1D-SM-stat}
\end{equation}%
cf. (\ref{1D-DNLS-stat}). Discrete solitons in the nonintegrable SM equation
(\ref{SAmodel}) with $\mu >0$, i.e. with \emph{noncompeting} intersite and
onsite self-focusing nonlinearities, were investigated by means of numerical
methods \cite{Cai, Cai97, Dmitriev03}. It has been demonstrated that the SM\
gives rise to static (and, sometimes, approximately mobile \cite{Cai97})
solitons at all positive values of $\mu $.

\begin{figure}[tb]
\begin{center}
\begin{tabular}{cc}
\includegraphics[width=5.5cm,scale=1]{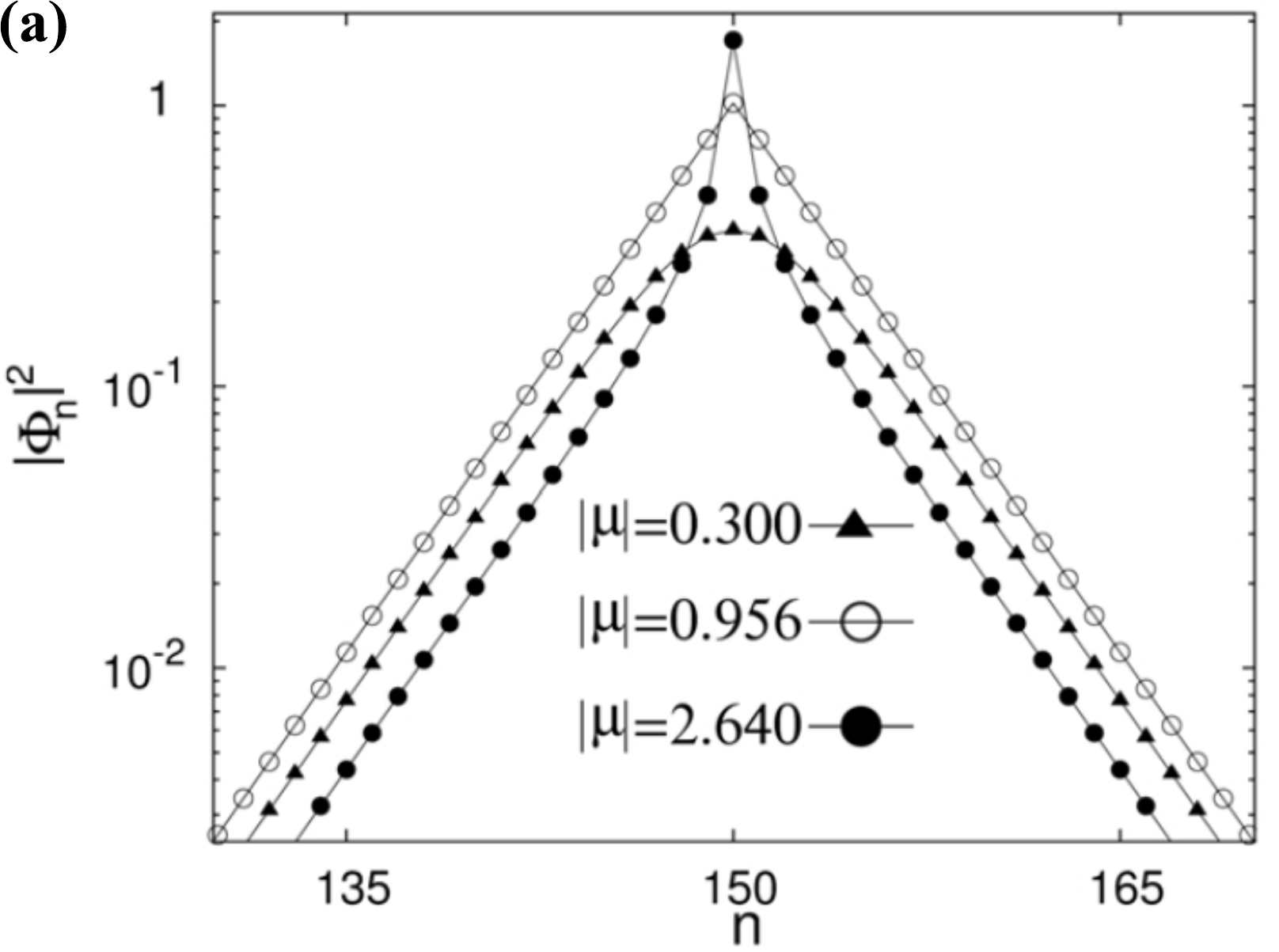} & %
\includegraphics[width=5.5cm,scale=1]{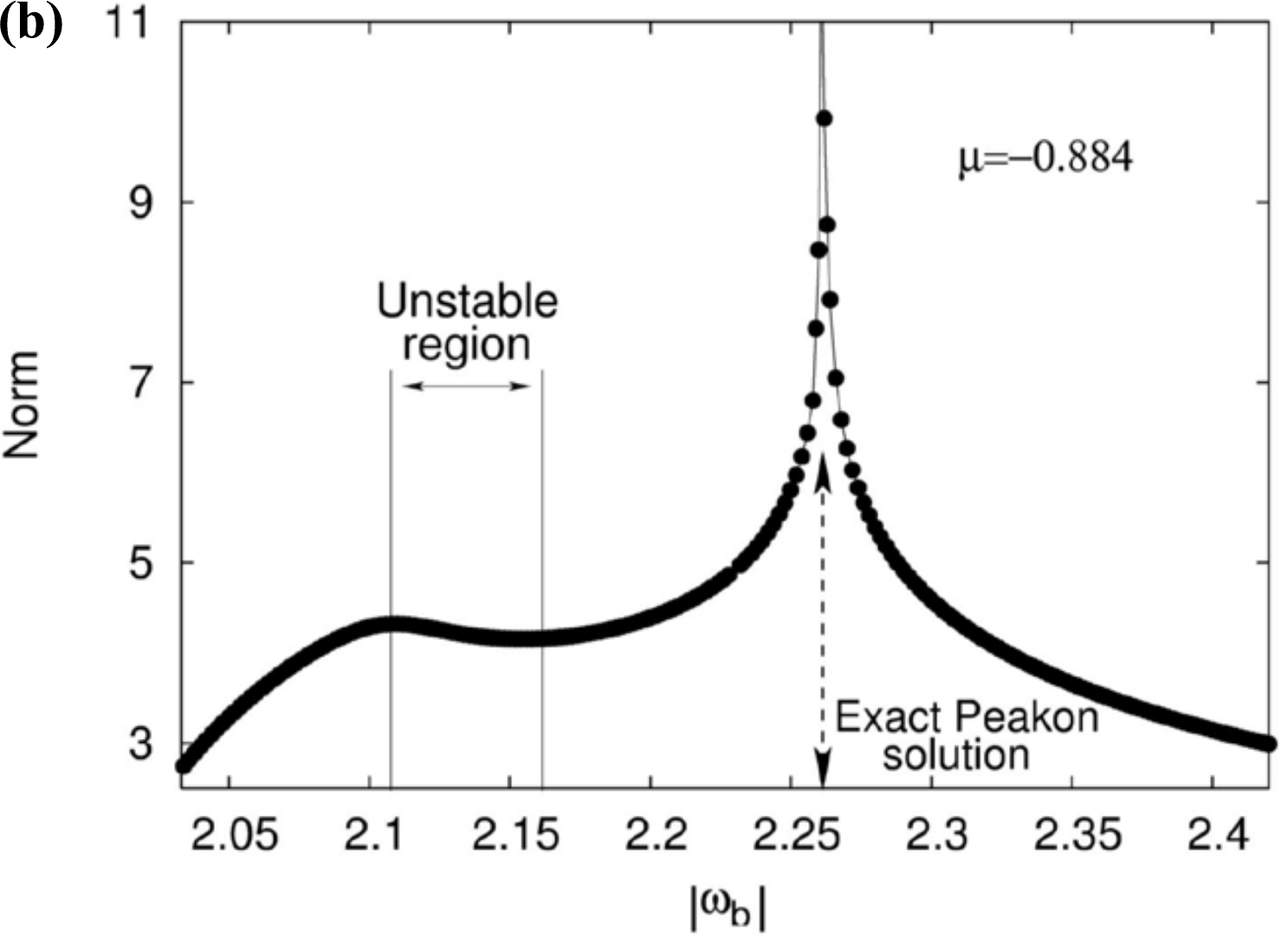}%
\end{tabular}%
\end{center}
\caption{(a) Examples of three different types of discrete solitons, shown
on the logarithmic scale, which are produced by Eq. (\protect\ref{1D-SM-stat}%
), i.e., the Salerno model, with competing nonlinearities ($\protect\mu <0$%
), at $\protect\omega =-2.091$: an ordinary soliton for $\protect\mu =-0.3$,
a peakon for $\protect\mu =-0.956$, and a cuspon for $\protect\mu =-2.64$.
In the figure, $\left\vert \Phi _{n}\right\vert $ has the same meaning as $%
u_{n}$ in Eq. (\protect\ref{1D-SM-stat}). (b) The norm of the discrete
solitons in the Salerno model with competing nonlinearities vs.\ the
frequency (denoted here $\protect\omega _{b}$, instead of $\protect\omega $%
), for $\protect\mu =-0.884$. Reproduced from \protect\cite{Zaragoza}.}
\label{fig8}
\end{figure}

Another possibility is to consider the SM with $\mu <0$, which features
\emph{competing nonlinearities}, as the intersite cubic term, with
coefficient $\mu <0$ in (\ref{SAmodel}), which accounts for nonlinear
coupling between adjacent sites of the lattice, and the onsite term in (\ref%
{SAmodel}) (the last term in that equation) represent, respectively,
self-defocusing and focusing nonlinear interactions. It was found \cite%
{Zaragoza} that this version of the SM gives rise to families of quiescent
discrete solitons, which are looked for in the usual form (\ref{psi-u}),
with $\omega <0$ and real amplitudes $u_{n}$, of two different types.\ One
family represents ordinary discrete solitons, similar to those generated by
the DNLS equation. Another family represents \textit{cuspons, }featuring
higher curvature of their profile at the center than exponential shapes.
Examples of numerically found stable discrete solitons of these types are
displayed in Fig. \ref{fig8}a. The border between the ordinary discrete
solitons and cuspons is represented by a special discrete mode, in the form
of a stable \textit{peakon}, which is also shown in Fig.~\ref{fig8}a.

The continuum limit of the SM with competing nonlinearities, given by (\ref%
{SAcont}) with $\mu <0$, produces continuous solitons in the usual form, $%
\Psi =\exp \left( -i\omega t\right) U(x)$, in the frequency band $0<-\omega
<1/|\mu |-1$, provided that $|\mu |<1$. At the edge of the soliton band,
i.e. at $\omega =-\left( 1/|\mu |-1\right) $, (\ref{SAcont}) gives rise to
an exact solution in the form of the continuous peakon 
$U_{\mathrm{peakon}}(x)=\left( 1/\sqrt{|\mu |}\right) \exp \left( -\sqrt{%
\left( 1/|\mu |\right) -1}|x|\right) $
\cite{Zaragoza}. For continuous solutions, the name of \textquotedblleft
peakon\textquotedblright\ implies a jump of the derivative at the central
point, 
while cuspons do not exist in the continuum limit.


The stability analysis of discrete solitons produced by the SM with
competing nonlinearities demonstrate that only a small subfamily of ordinary
solitons is unstable, while all cuspons, including the peakon, are stable.
For fixed $\mu =-0.884$, a typical situation for families of discrete
solitons in the SM with competing nonlinearities is presented in Fig.~\ref%
{fig8}b, which shows norm (\ref{eq:Norm}) as a function of $|\omega |$. The
plot clearly demonstrates that ordinary solitons and cuspons are separated
by the peakon, as mentioned above. Except for the part of the
ordinary-soliton family with the negative slope, $d{N}/d(|\omega |)<0$,
which is marked in Fig.~\ref{fig8}b, the discrete solitons are stable. In
particular, it is worthy to note that the cuspons and peakon are completely
stable modes. The instability of the segment of the family of ordinary
discrete solitons with $d{N}/d(|\omega |)<0$ exactly agrees with the
prediction of the well-known Vakhitov-Kolokolov criterion \cite{Vakh}. On
the other hand, it is seen from Fig. \ref{fig8}b that the VK criterion,
being valid for the ordinary solitons, is actually \emph{reversed} for the
cuspons \cite{Zaragoza}. 


As mentioned above, antisymmetric bound states of DNLS solitons are stable,
while symmetric bound states are unstable \cite{bound states,bound states 2}%
. As shown in \cite{Zaragoza}, the same is true for bound states of ordinary
discrete solitons in the SM. However, a noteworthy finding is that, in the
framework of the SM with competing nonlinearities, the situation is \emph{%
exactly opposite} for the cuspons: their symmetric and antisymmetric bound
states are stable and unstable, respectively \cite{Zaragoza}.

\subsection{The two-dimensional Salerno model and its discrete solitons}

The 2D version of the SM was introduced in \cite{Zaragoza2D}. It is based on
the following equation, cf. (\ref{SAmodel}),
\begin{eqnarray}
i\dot{\psi}_{n,m} &=&-\left[ \left( \psi _{n+1,m}+\psi _{n-1,m}\right)
+C\left( \psi _{n,m+1}+\psi _{n,m-1}\right) \right]  \notag \\
&\times &\left( 1+\mu \left\vert \psi _{n,m}\right\vert ^{2}\right)
-2\left\vert \psi _{n,m}\right\vert ^{2}\psi _{n,m}\;,  \label{2dSalerno}
\end{eqnarray}%
%
%
%
%
%
%
%
%
%
%
%
%
%
%
where real constant $C>0$ accounts for a possible anisotropy of the 2D
lattice. Similar to its 1D version, Eq. (\ref{2dSalerno}) conserves the norm
and Hamiltonian, cf. (\ref{eq:Norm}) and (\ref{HSM}),
\begin{equation}
\left( N_{\mathrm{AL}}\right) _{\mathrm{2D}}=(1/\mu )\sum_{m,n}\ln
\left\vert 1+\mu |\psi _{n,m}|^{2}\right\vert \;,  \label{SalernoNorm}
\end{equation}%
\begin{gather}
\left( H_{\mathrm{AL}}\right) _{\mathrm{2D}}=-\sum_{n,m}\left[ \left( \psi
_{n,m}\psi _{n+1,m}^{\ast }+\psi _{n+1,m}\psi _{n,m}^{\ast }\right) +C\left(
\psi _{n,m}\psi _{n,m+1}^{\ast }+\psi _{n,m+1}\psi _{n,m}^{\ast }\right)
\right.  \notag \\
\left. +(2/\mu )|\psi _{n,m}|^{2}\right] +(2/\mu )\left( N_{\mathrm{AL}%
}\right) _{\mathrm{2D}}.\;  \label{eq:SalernoHam}
\end{gather}%
The continuum limit of this model is a 2D continuous equation which is a
straightforward extension of its 1D counterpart (\ref{SAcont}):%
\begin{equation}
i\Psi _{t}+\left( 1+\mu \left\vert \Psi \right\vert ^{2}\right) \left( \Psi
_{xx}+\Psi _{yy}\right) +2\left[ (1+C)\mu +1\right] |\Psi |^{2}\Psi =0.
\label{cont}
\end{equation}%
%
%
%
%
Note that term in $\mu \left\vert \Psi \right\vert ^{2}\left( \Psi
_{xx}+\Psi _{yy}\right) $ prevents the onset of the collapse in Eq.~(\ref%
{cont}).

\begin{figure}[tb]
\begin{center}
\includegraphics[width=11cm,scale=1,clip]{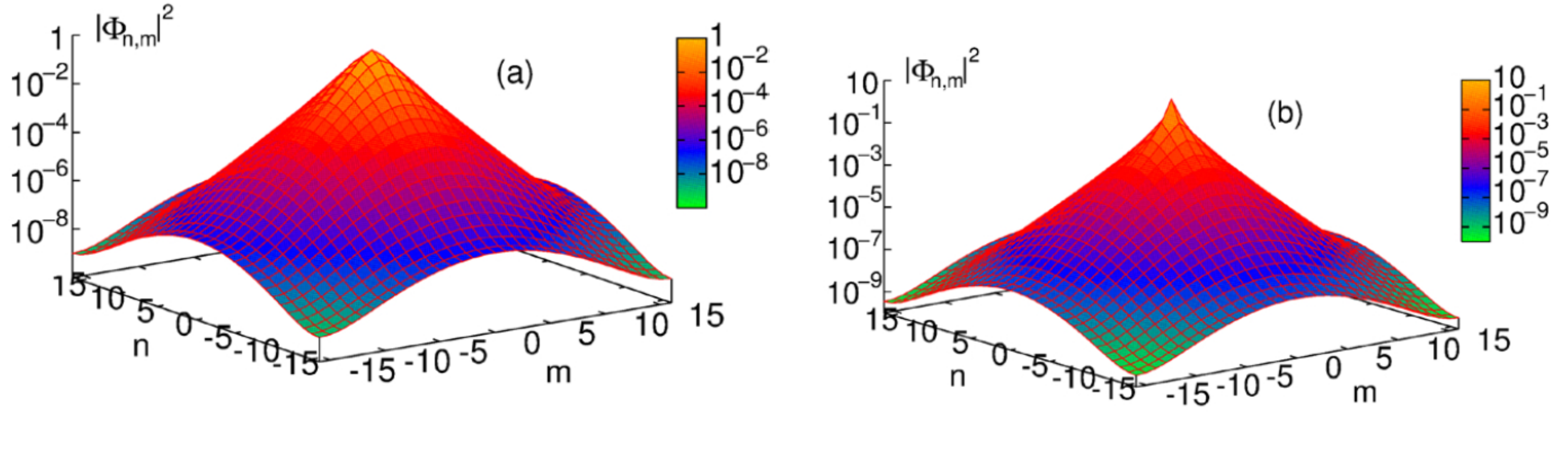}
\end{center}
\caption{Discrete solitons in the isotropic ($C=1$) 2D Salerno model with
competing nonlinearities [$\protect\mu <0$ in (\protect\ref{2dSalerno})],
obtained for frequency $\protect\omega =-4.22$: (a) a regular soliton at $%
\protect\mu =-0.2$; \textbf{(}b\textbf{)} a cuspon at $\protect\mu =-0.88$.
Reproduced from \protect\cite{Zaragoza2D}.}
\label{fig10}
\end{figure}

2D solitons are looked for in the same form as their 1D counterparts, $\psi
_{mn}(t)=e^{-i\omega t}u_{mn}$, cf. (\ref{psi-u}). In the most interesting
case of the competing nonlinearities, $\mu <0$, the situation is similar to
that outlined above for SM in 1D: there are ordinary discrete solitons,
which have their stability and instability regions, and 2D cuspons, which
are \emph{entirely stable} in their existence region. Typical 2D solitons of
both types are displayed in Fig.~\ref{fig10}. Also similar to the 1D case,
ordinary solitons and cuspons are separated by 2D peakons, which are stable.
In addition to that, antisymmetric bound states of ordinary 2D discrete
solitons, and symmetric complexes built of 2D cuspons, are stable, while the
bound states with opposite parities are unstable, also like in the 1D model.

Along with the fundamental solitons, the 2D SM with the competing
nonlinearities gives rise to vortex-soliton modes with narrow stability
regions \cite{Zaragoza2D}. In the 2D SM with non-competing nonlinearities,
unstable vortex solitons spontaneously transform into fundamental solitons,
losing their vorticity (this is possible because the angular momentum is not
conserved in the lattice system). The situation is essentially different in
the 2D SM with competing nonlinearities, where unstable vortex modes
transform into \textit{vortical breathers}, i.e., persistently oscillating
localized modes that keep the original vorticity.

\section{A brief survey of semi-discrete systems}


A topic which may be a subject for a separate review, is semi-discrete
systems, i.e., 2D settings which are discrete in one direction and
continuous in the other. Accordingly, such systems can create semi-discrete
solitons. A system of this type which was explored in detail is an array of
optical fibers \cite{Rubenchik}, modeled by a system of coupled NLS
equations for amplitudes $u_{n}\left( z,\tau \right) $ of electromagnetic
waves in individual fibers:
\begin{equation}
i\partial _{z}u_{n}+(1/2)D\partial _{\tau }^{2}u_{n}+(\kappa /2)\left(
u_{n+1}+u_{n-1}-2u_{n}\right) +\left\vert u_{n}\right\vert ^{2}u_{n}=0,
\label{kappa}
\end{equation}%
where $D$ is the group-velocity-dispersion coefficient in each fiber, and $%
\kappa >0$ is the coefficient of coupling between adjacent fibers in the
array. It supports semi-discrete solitons in the case of anomalous
dispersion, i.e. $D>0$. A remarkable property of semi-discrete modes
generated by Eq. (\ref{kappa}) is their ability to stably move \emph{across
the array}, under the action of a \textit{kick} applied to them at $z=0$
\cite{Blit}:%
\begin{equation}
u_{n}(\tau )\rightarrow \exp \left( ian\right) u_{n}(\tau ),  \label{kick}
\end{equation}%
with real $a$. An example of such a moving mode is displayed in Fig. \ref%
{fig11}. This property may be compared to the above-mentioned mobility of 1D
discrete solitons in the DNLS equation \cite{Feddersen}.
\begin{figure}[t]
\includegraphics[width=6cm,scale=1]{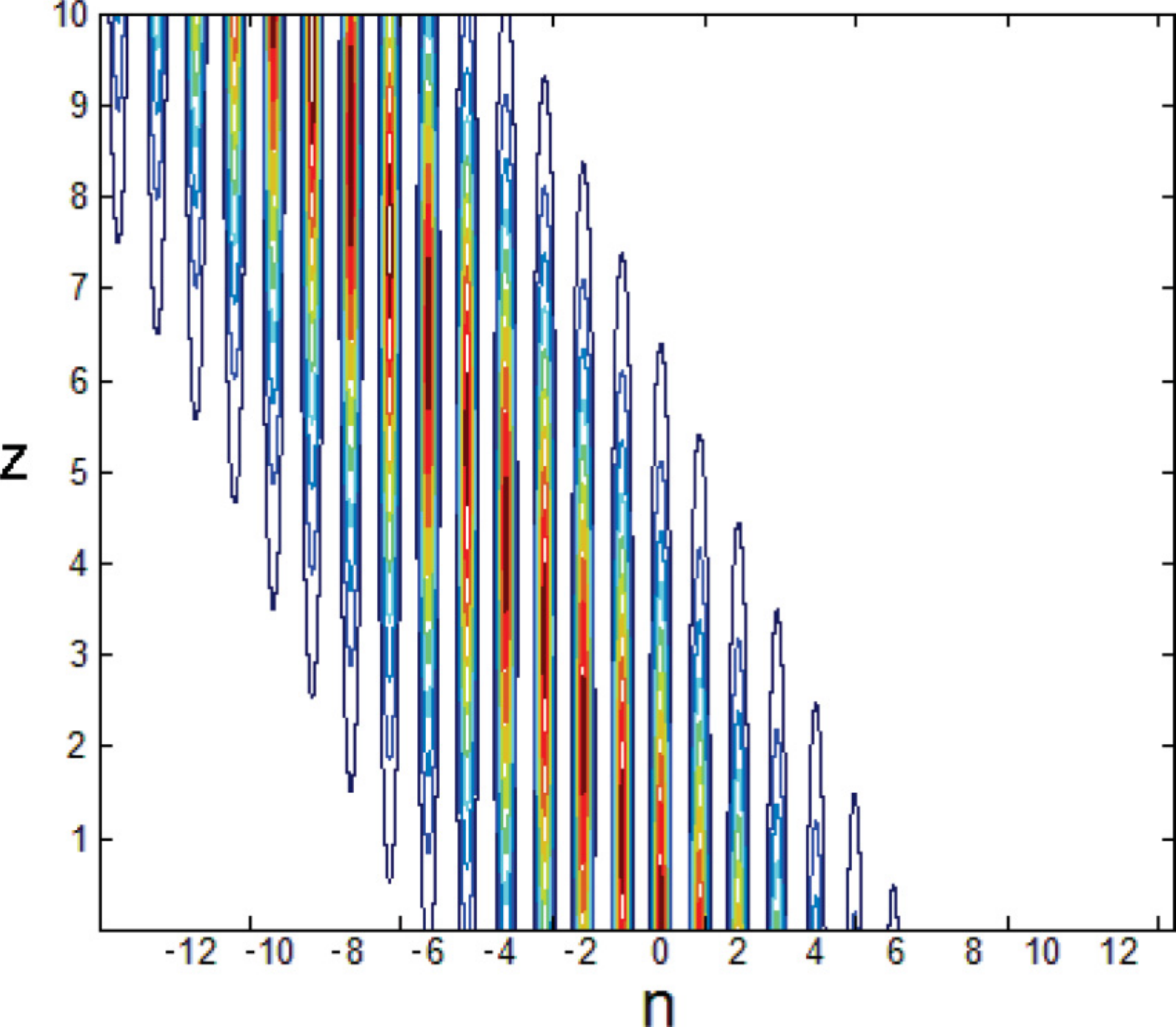} 
\caption{An example of a semi-discrete spatiotemporal mode, generated by Eq.
(\protect\ref{kappa}), which performs stable transverse motion under the
action of the kick, defined according to (\protect\ref{kick}), with $a=1.5$.
The cross section of the plot at any fixed $z$ shows the distribution of
power $\left\vert u_{n}(\protect\tau )\right\vert ^{2}$ for each $n$.
Reproduced from \protect\cite{Blit}.}
\label{fig11}
\end{figure}

\begin{figure}[tb]
\begin{center}
\includegraphics[width=12cm,scale=1]{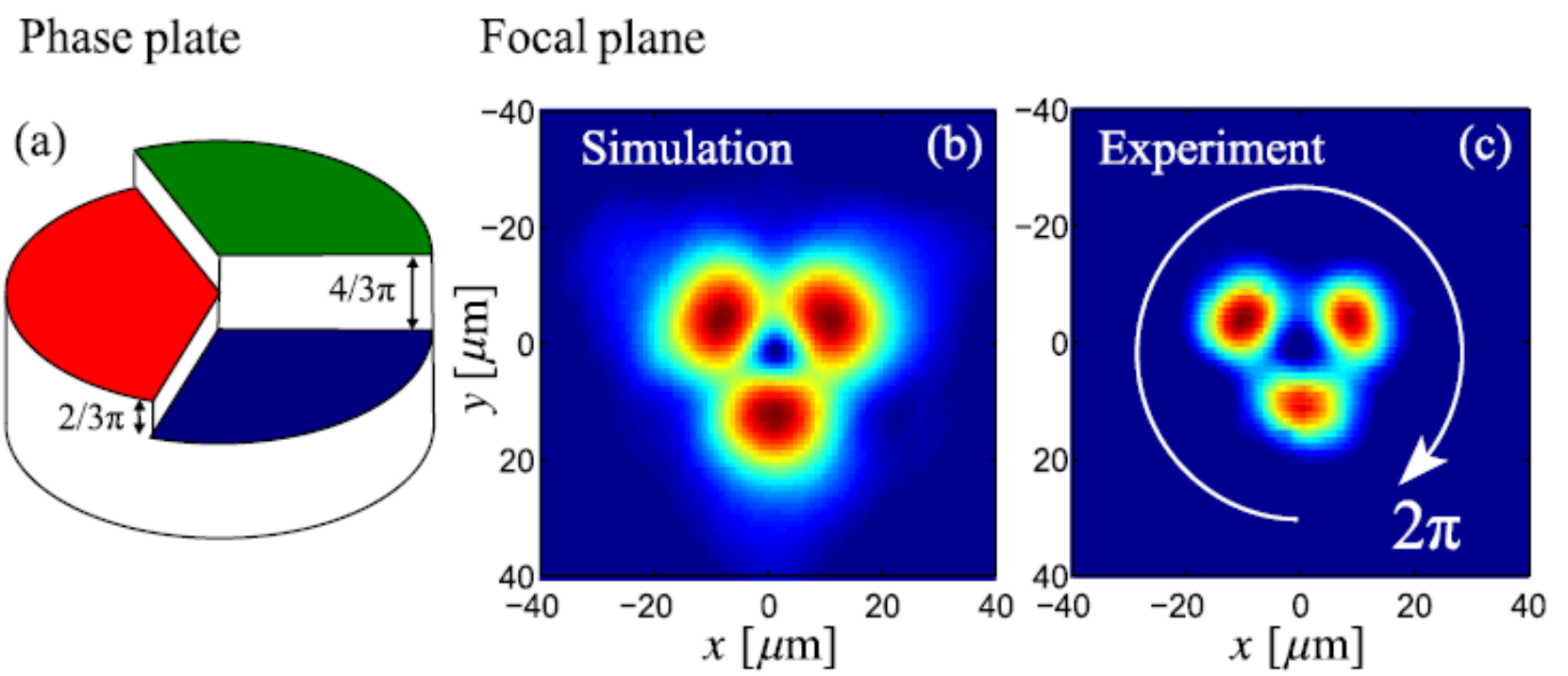}
\end{center}
\caption{A semi-discrete vortex soliton in a hexagonal quasi-discrete array
of waveguides made in bulk silica. (a) A phase plate used for imprinting the
vortex structure into the input beam. (b,c) Numerically simulated and
experimentally observed (transient) intensity distributions in the
transverse plane, with phase shifts $2\protect\pi /3$ between adjacent
peaks. Reproduced from \protect\cite{Jena-vort}. Creative Commons
Attribution License (CC BY) \url{http://creativecommons.org/licenses/by/3.0/}%
.}
\label{fig12}
\end{figure}

Similarly, quasi-discrete settings modeled by an extension of (\ref{kappa})
with two transverse spatial coordinates were used for the creation for
spatiotemporal optical solitons (``light bullets") \cite{Jena}, as well as
soliton-like transient modes with embedded vorticity \cite{Jena-vort}.
Waveguides employed in those experiments feature a transverse
hexagonal-lattice structure, written in bulk silica by means of an optical
technology. A spatiotemporal\ vortex state (in the experiment, it is
actually a transient one) in the bundle-like structure is presented by Fig. %
\ref{fig12}, which displays both numerically predicted and experimentally
observed distributions of intensity of light in the transverse plane,
together with a phase plate used in the experiment to embed the vorticity
into the incident spatiotemporal pulse which was used to create the mode.

\begin{figure}[tb]
\begin{center}
\includegraphics[width=12cm,scale=1]{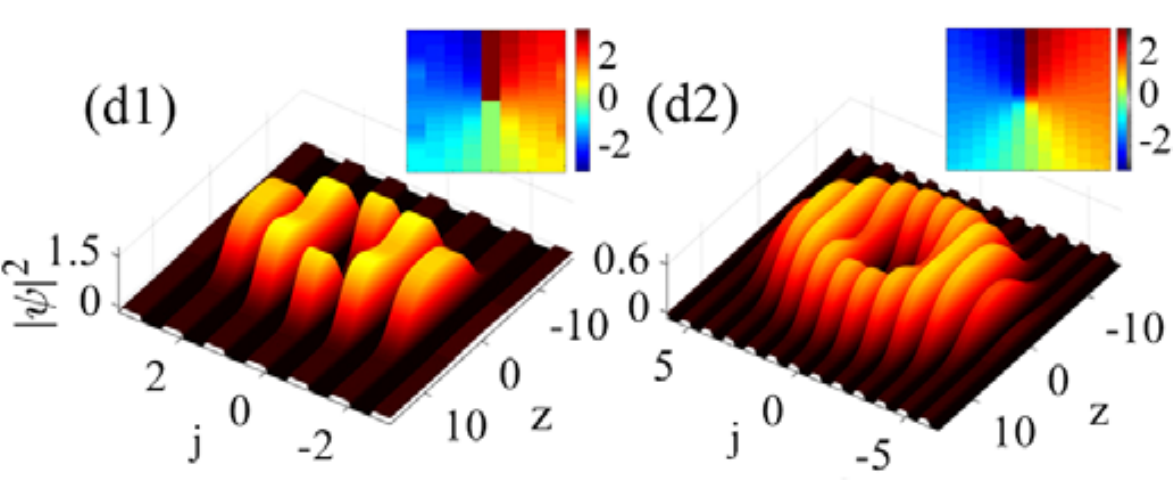}
\end{center}
\caption{Left and right panels display, respectively, examples of amplitude
and phase profiles of stable OC and IC semi-discrete vortex solitons
produced by Eq. (\protect\ref{GPE}). Reproduced from \protect\cite{Raymond}.}
\label{fig13}
\end{figure}

A new type of semi-discrete solitons was recently reported in \cite{Raymond}%
, in the framework of an array of linearly coupled 1D GPEs, including the
\textit{Lee-Hung-Yang correction}, which represents an effect of quantum
fluctuations around the mean-field states of a binary BEC \cite%
{Petrov1,Petrov2}. The system is%
\begin{equation}
i\partial _{t}\psi _{j}=-{(1/2)}\partial _{zz}\psi _{j}-\left( 1/2\right)
\left( \psi _{j+1}-2\psi _{j}+\psi _{j-1}\right) +g|\psi _{j}|^{2}\psi
_{j}-|\psi _{j}|\psi _{j},  \label{GPE}
\end{equation}%
where $\psi _{j}(z)$ is the mean-field wave function in the $j$-th core, the
self-attractive quadratic term represents the Lee-Hung-Yang effect, and $g>0$
accounts for the mean-field self-repulsion. This system gives rise to many
families of semi-discrete solitons, including a novel species of \textit{%
semi-discrete vortex solitons}. Typical examples of such stable states are
displayed in Fig. \ref{fig13}.


Semi-discreteness of another type is possible in two-component systems,
where one component is governed by a discrete equation, while the other one
obeys a continuous equation. This type of two-component systems was
introduced in \cite{Panoiu}, addressing a second-harmonic-generating model,
assuming that the continuous second-harmonic wave
propagates in a slab with a continuous transverse coordinate, 
while the fundamental-harmonic field 
is concentrated in a discrete waveguiding array attached to the slab.
Semi-discrete solitons of an inverted type, with the continuous
fundamental-frequency component and discrete second harmonic one, were also
constructed in \cite{Panoiu}. 

\section{Conclusion}

\subsection{Summary of the Chapter}

The interplay of discreteness and intrinsic nonlinearity in various physical
media gives rise to a great variety of static and dynamical states. Among
them, especially interesting are self-trapped ones in the form of discrete
solitons. The present Chapter aims to briefly review basic theoretical
models combining discreteness and nonlinearity, and basic results for
discrete solitons produced by such models. Essential experimental findings
are included too (in particular, those for 2D and 3D discrete solitons with
embedded vorticity). In many cases, discreteness helps to produce states
which either do not exist or are unstable in continuum counterparts of
discrete settings. In particular, the 1D DNLS equation gives rise to stable
bound states of fundamental solitons, and the 2D DNLS equation readily
creates fundamental and vortex solitons, whose counterparts are completely
unstable in the continuum. On the other hand, some properties which are
obvious in the continuum limit, such as mobility of solitons, are
problematic in discrete settings.

The work in this area is currently in progress, and new results may be
expected. A promising direction is to generate discrete counterparts of
complex continuous modes with intrinsic topological structures. Some results
obtained in this direction have already been reported, such as discrete
solitons in a system with spin-orbit coupling \cite{Sandra}, sophisticated
3D discrete modes with embedded vorticity \cite{3Dvort,3Dvort2}, and
discrete skyrmions \cite{skyrmion}. A challenging task is experimental
realization of such results which, thus far, were only predicted in the
theoretical form.

\subsection{Topics not included in the Chapter}

Due to length limitations, some essential models and methods are not
considered here. One of them is the \textit{anti-continuum limit}, which
makes it possible to obtain \textquotedblleft stems\textquotedblright\ for
many families of discrete solitons by considering, at first, lattice models
with no coupling between the sites. Using this approach, one can construct a
great deal of modes, by formally putting together various solutions
supported by non-interacting sites of the lattice. Then, the analysis allows
one to identify solution branches that can be extended to small nonzero
values of the intersite coupling. This method is efficient in constructing
many families of discrete solitons in diverse models \cite{Aubry1,bound
states 2,they,Aubry2}.

Interaction of discrete solitons with local defects in the underlying
lattice, as well as with interfaces and edges (surfaces, if the underlying
lattice is two- or three-dimensional) is another vast area of theoretical
and experimental studies. In particular, defects and surfaces may often help
to create and stabilize localized modes which do not exist or are unstable
in uniform lattices, such as Tamm \cite{Tamm} and topological-insulator \cite%
{top-ins,top-ins2} states.

Large topics are solitons in discrete dissipative nonlinear systems, and in
systems subject to the condition of the parity-time ($\mathcal{PT}$)
symmetry. In this Chapter, dissipative systems, which include friction and
driving forces, are considered only in terms of TL and FK models (in
particular, for arrays of Josephson junctions, in Fig. \ref{fig1}). Other
dissipative versions of TL\ models are known too, in the form of LC
transmission lines for electric pulses. They support traveling discrete
solitons, which have been produced in theoretical and experimental forms
\cite{LC1,LC2,LC3}.

In other contexts, basic nonlinear dissipative models are represented by
discrete complex Ginzburg-Landau equations, i.e., DNLS equations with
complex coefficients in front of onsite linear and nonlinear terms, which
account for dissipative losses and compensating gain \cite{Hakim}. These
models give rise to discrete solitons which do not exist in continuous
families, unlike the DNLS solitons, but rather as isolated \textit{attractors%
} \cite{Efremidis1,Akhmed,Efremidis2}.

Systems with $\mathcal{PT}$ symmetry are dissipative models which share many
properties with conservative ones. They include mutually symmetric spatially
separated linear gain and loss elements \cite{Bender,Christod,Christod2}.
This arrangement makes it natural to consider $\mathcal{PT}$-symmetric
systems with a discrete structure. Their experimental realization in optics
\cite{Christod2} suggests to include the Kerr nonlinearity, thus opening the
way to prediction of $\mathcal{PT}$-symmetric discrete solitons \cite%
{PTrev1,PTrev2}. In particular, various species of stable 1D and 2D discrete
solitons were predicted in chains of $\mathcal{PT}$-symmetric elements \cite%
{PTsol1,PTsol0,PTsol2,PTsol3,PTsol4,PTsol5}. An example of $\mathcal{PT}$%
-symmetric solitons has been created experimentally in a similar discrete
setting \cite{PTsol-observation}.

\section*{Acknowledgements}
I appreciate the invitation of Editors of volume \emph{Nonlinear
Science: a 20/20 vision} to submit this Chapter. One of the Editors, Prof. J. Cuevas-Maraver,
has provided a great deal of help in the course of preparing the manuscript.
I would like to thank colleagues in collaboration with whom I have been working
on various topics related to the review: G.E. Astrakharchik,
P. Beli\v{c}ev, A.R. Bishop, 
L.L. Bonilla, R. Carretero-Gonz\'{a}lez, Zhaopin Chen, Zhigang Chen, C. Chong,
M. Cirillo, J. Cuevas-Maraver, J.
D'Ambroise, F.K. Diakonos, S.V. Dmitriev, 
L.M. Flor\'{\i}a, D.J. Frantzeskakis, S. Fu, G. Gligori\'{c},
J. G\'{o}mez-Garde\~{n}es, N. Gr\o nbech-Jensen, L. Had\v{z}ievsli, D. Herring,
J. Hietarinta, 
N.V. Hung, Y.V. Kartashov, T. Kapitula, D.J. Kaup,
P.G. Kevrekidis, V.V. Konotop, T. Kuusela, M. Lewenstein, Y. Li, A.
Maluckov, 
N.C. Panoiu, I.E. Papachalarampus, 
M.A. Porter, K.\O. Rasmussen, H. Sakaguchi, L. Torner, M. Trippenbach,
A.V. Ustinov, R.A. Van Gorder, M.I. Weinstein. 


\end{document}